\documentclass[journal]{IEEEtran}
\newcommand{\RNum}[1]{\uppercase\expandafter{\romannumeral#1\relax}}
\usepackage{diagbox}
\usepackage{graphicx}
\usepackage{url}
\usepackage{hyperref}
\usepackage{gensymb}
\usepackage{amsmath}
\usepackage{float}
\usepackage{booktabs}
\usepackage{epstopdf}
\usepackage[justification=centering]{caption}
\usepackage{cite}
\usepackage{amsfonts}
\usepackage{tabularx}
\usepackage{booktabs}
\usepackage{threeparttable}
\usepackage{upgreek}
\usepackage{subfigure}
\usepackage{color,soul}
\usepackage{mathtools}
\usepackage{multirow}
\usepackage[figuresright]{rotating}
\usepackage{adjustbox}
\usepackage{textcomp}
\usepackage{amssymb}
\usepackage{ulem}

\usepackage{amsthm}
\usepackage{stmaryrd}
\usepackage{dsfont}
\usepackage{xfrac}
\usepackage{mathrsfs}

\usepackage[ruled]{algorithm2e}

\soulregister\cite7
\soulregister\citep7
\soulregister\citet7
\soulregister\ref7
\soulregister\pageref7

\begin{document}

	\title{Code-aided CFOs and CPOs Estimation in Cooperative Satellite Communication}
	%Multiple Satellites Collaboration for Joint Code-aided CFOs and CPOs Estimation
		\author{
		Pingyue Yue,~\IEEEmembership{Student Member,~IEEE}, Yixuan Li,~\IEEEmembership{Student Member,~IEEE}, Yue Li, \\
		Rui Zhang,~\IEEEmembership{Member,~IEEE}, Shuai Wang,~\IEEEmembership{Member,~IEEE}, and Jianping An~\IEEEmembership{Member,~IEEE}
		
		\thanks{
%			 \emph{\textbf{(Corresponding author)}}			
						Pingyue Yue and  Yixuan Li are with the School of Information and Electronics, Beijing Institute of Technology, Beijing 100081, China (e-mails: ypy@bit.edu.cn; ethanlee@bit.edu.cn). 
						
						Yue Li is with the System Demonstration and Research Institute of Navy Academy of Equipment, Shanghai, China (e-mails:li$\_$yue98@hotmail.com). 
						
						Rui Zhang, Shuai Wang, and Jianping An are with the School of Cyberspace Science and Technology, Beijing Institute of Technology, Beijing 100081, China (e-mails: rui.zhang@bit.edu.cn; swang@bit.edu.cn; an@bit.edu.cn). 
						
		}
	}
	\maketitle
	
	\begin{abstract}		
		Low Earth Orbit (LEO) satellites have gained significant research attention in the development of secure Internet of Remote Things (IoRT). 
        In scenarios where miniaturized terminals are involved, the limited transmission power and long transmission distance often result in a low Signal-to-Noise Ratio (SNR) at the satellite receiver, leading to degraded communication performance. 
        To mitigate this issue, the use of cooperative satellites has been proposed, which can combine signals received from multiple satellites, thereby improving the SNR considerably. However, achieving the maximum combination gain requires synchronization of carrier frequency and phase for each receiving signal, which poses a challenge under low SNR conditions, particularly in short burst transmissions without training sequences.
        To address this challenge, we propose an iterative code-aided estimation algorithm for joint estimation of Carrier Frequency Offset (CFO) and Carrier Phase Offset (CPO). The algorithm incorporates a two-step estimation procedure that utilizes Iterative Cross Entropy (ICE) and Cooperative Expectation Maximization (CEM). The ICE method is employed initially to perform a coarse search for parameter estimation, followed by the CEM technique which refines the estimates. The performance limit of parameter estimation is evaluated using the Cramér-Rao Lower Bound (CRLB).
        Simulation results indicate that the proposed algorithm achieves estimation accuracy close to the CRLB within the frequency range of ($-7.8125\times10^{-3}$, $+7.8125\times10^{-3}$] and phase range of ($-\pi $, $+\pi$]. Furthermore, the algorithm demonstrates the ability to approach the Bit Error Rate (BER) performance bounds, with deviations of $0.3$ dB and $0.4$ dB in scenarios involving two-satellite and four-satellite collaboration, respectively.
		
		%这部分最后是在两颗星和四颗星场景下分别有0.3和0.4dB的损失
		
	\end{abstract}
	
	\begin{IEEEkeywords}
		Cooperative satellite communication, low signal to noise ratio, short burst transmission, signal coherent combining, carrier frequency offset, carrier phase offset.
	\end{IEEEkeywords}

	\section{Introduction}
	
	Internet of Remote Things (IoRT) is a network of small objects which are often dispersed over wide geographical areas even inaccessible. 
	In IoRT, satellite communication can provide a cost-effective solution to their interconnection and communication in comparison to terrestrial networks \cite{Pingyue1,Pingyue2,IoRT2016}. 
	However, due to the limited link budgets caused by the constrained transmission power and long distance between the satellite and terminal, the satellite receiver has to operate at low Signal-to-Noise Ratio (SNR). 
	The reception and processing of low SNR signals has traditionally been a challenging task in signal processing. One attractive approach to addressing this challenge is the use of cooperative diversity techniques. In recent years, Cooperative Satellite Communication (CSC) has garnered significant attention due to its potential to enhance communication reliability and efficiency in satellite networks \cite{Arapoglou2011MIMOOS}. This technique is particularly useful for remote or energy-limited terminals, such as those employed in IoRT. With the rapid expansion of Low Earth Orbit (LEO) constellations, it has become feasible to implement collaboration between multiple satellites. Consequently, multi-satellite signal combining can be an effective strategy to achieve efficient spectral resource utilization and improve communication performance in low SNR environment.	
	
	In the context of IoRT systems utilizing CSC, uplink cooperation can facilitate the collaborative reception of signals from a single terminal by multiple satellites. This enables effective signal combination through the integration of received signals. However, due to factors such as varying distances between satellites and terminals, antenna directivity, relative velocities, and crystal oscillator drift on board the satellites, it is possible that the signal amplitude, propagation delay, Carrier Frequency Offset (CFO), and Carrier Phase Offsets (CPO) may vary significantly among satellites. If these parameters are not aligned, diversity reception performance may deteriorate, leading to infeasible demodulation. To address this issue, accurate estimation of parameter differences among received signals and optimal combining weights is crucial for successful signal combination. 
	
	The current studies on the estimation of CFO and CPO have primarily focused on individual reception scenarios. These estimation methods can be broadly categorized into two approaches: Data-Aided (DA) and Non-Data-Aided (NDA). DA parameter estimation requires the use of a known pilot symbol sequence as a training sequence, which may limit its applicability and reduce the effective data rate and spectral efficiency of the system \cite{CFEstimation2000}. In contrast, NDA estimation directly estimates the parameters based on signal characteristics without the need for additional known data. However, NDA methods can introduce demodulation noise that degrades synchronization accuracy, especially in low SNR scenarios \cite{Mohammadkarimi2016NDA}. Recent studies have focused on Code-Aided (CA) carrier synchronization algorithms that utilize the decoding results from the decoder to estimate carrier parameters \cite{Chaofan1, bellili2015closed}. This innovative approach incorporates coding gain into the carrier synchronization process, resulting in improved performance, particularly in low SNR scenarios where pilot symbols are not needed. The CA technique has demonstrated effectiveness, especially in short-burst communication systems \cite{Rahamim2008}, making it a promising approach for IoRT applications.

    However, the CA synchronization method has its own limitations. Firstly, it is constrained by significant limitations on the estimation range of both CFO and CPO. As explained in Section III of this paper, to achieve a decoding performance with a Bit Error Rate (BER) below 0.5, the Normalized Frequency Offset (NFO) (which represents the CFO normalized to the symbol rate) must be smaller than 1e-4, while the CPO must fall within the range of $[-0.2\pi,0.2\pi]$. Consequently, the applicability of the CA method is inherently restricted in practical scenarios. In an effort to overcome this challenge, previous research proposes a two-dimensional exhaustive search within a narrow CFO range while the CPO is in $(-\pi,\pi]$ has been proposed in literature \cite{NovelCA2013}. This is coupled with an interpolation algorithm to increase estimation accuracy during the search process. However, due to noise introduced during the single-step search and interpolation processes, as well as the high dependence of algorithm performance on predefined thresholds and initial search point selection, there may  exist performance degradation at low SNR levels. Additionally, an algorithm presented in \cite{Iterative2020} strives to extend the CFO estimation range. It leverages a coarse search and a fine search with code assistance to achieve a refined estimation of the CFO. However, this Gaussian process remains sensitive to the CFO as it is derived from decoding results. Secondly, the CA method exhibits reduced decoding reliability at low SNR levels, leading to a higher BER in the recovery of the modulation information from the transmitted data. This weakens estimation accuracy and highlights the sensitivity to CFO and CPO and the reliability of its decoding output. Despite the coding gain benefits provided by the CA synchronization method, its estimation accuracy is limited by residual frequency bias and random phase bias.

     with regards to CSC, existing research primarily focuses on mitigating jamming \cite{LEOdiversity}, capacity analysis \cite{LEOMIMOCap1, AccessMIMO, TcomDeng}, and capacity optimization \cite{MIMOOptim1}. However, these studies often assume that synchronization between satellites and terminals has already been achieved, which is not always the case in practical implementations. It is worth noting that there is a lack of comprehensive investigation into the estimation of CFO and CPO in the context of CSC. While previous work on CFO and CPO estimation in individual reception has shown improved performance using the CA method without a training sequence, this method is not suitable for scenarios with lower SNR levels and phase offsets within the range of $(-\pi, +\pi]$ in CSC.

     To overcome these limitations, we propose an iterative CA estimation algorithm comprised of two components: Iterative Cross Entropy (ICE) and Cooperative Expectation Maximization (CEM). ICE performs a coarse search for CFOs and CPOs, while CEM is responsible for fine estimation. Our proposed algorithm enables accurate estimation and compensation of CFOs and CPOs, as well as coherent combination of multiple satellites, even under challenging conditions with lower SNR levels and phase offsets within the range of $(-\pi, +\pi]$ in CSC scenarios. In summary, the contributions of this paper can be summarized as follows

		\begin{itemize}
			\item  We present a comprehensive estimation framework for CSC that allows for iterative estimation of CFOs and CPOs without the need for training sequences. It allows for precise estimation of CFOs and CPOs within large residual CFOs and random CPOs in the range of $(-\pi, +\pi]$. Furthermore, we leverage the combined decoding results during each iteration to enhance the estimation accuracy under lower SNR conditions for each satellite.
			
			\item We propose an iterative estimation algorithm based on cross entropy, enabling parallel joint estimation of CFOs and CPOs while accounting for the large residual CFOs and random CPOs within the range of $(-\pi, +\pi]$. During the coarse estimation process, we quantize the potential CFO range and utilize the corrected CFOs to compensate the received signals. Subsequently, we demodulate and square the obtained results to estimate the CPOs. By iteratively combining the signals after CFO and CPO correction in a quasi-coherent manner, using the combined SNR loss as the objective function, we achieve joint estimation of CFOs and CPOs for multiple satellites with only a few iterations.

            \item We propose a Maximum Likelihood Estimation (MLE) algorithm based on Expectation Maximization (EM) iteration to accurately estimate CFOs and CPOs under low SNR conditions. In each iteration, we compensate the received signal at each satellite using the estimated CFO and CPO to achieve coherent combination. The resulting combined decoding results are then utilized to assist in the following estimation of CFOs and CPOs. Through several iterations, we obtain accurate estimations of CFOs and CPOs for each satellite, as well as the coherent combined decoding results.
			
		\end{itemize}
		
  The remainder of this paper is organized as follows. 
		Section II outlines the system model for uplink CSC. 
		Section III discusses the problem formulation for joint CFOs and CPOs estimation in CSC.
		In section IV, the iterative estimation algorithm is proposed, which consists of the coarse estimation based on iterative cross entropy and fine estimation based on cooperative expectation maximization. 
		Simulation results and performance analysis are then shown in Section V. 
		Ultimately, Section VI draws the conclusions.
	
	\section{System Model}
	
	\begin{figure*}[ht]
		\centering
		\includegraphics[width=1\textwidth]{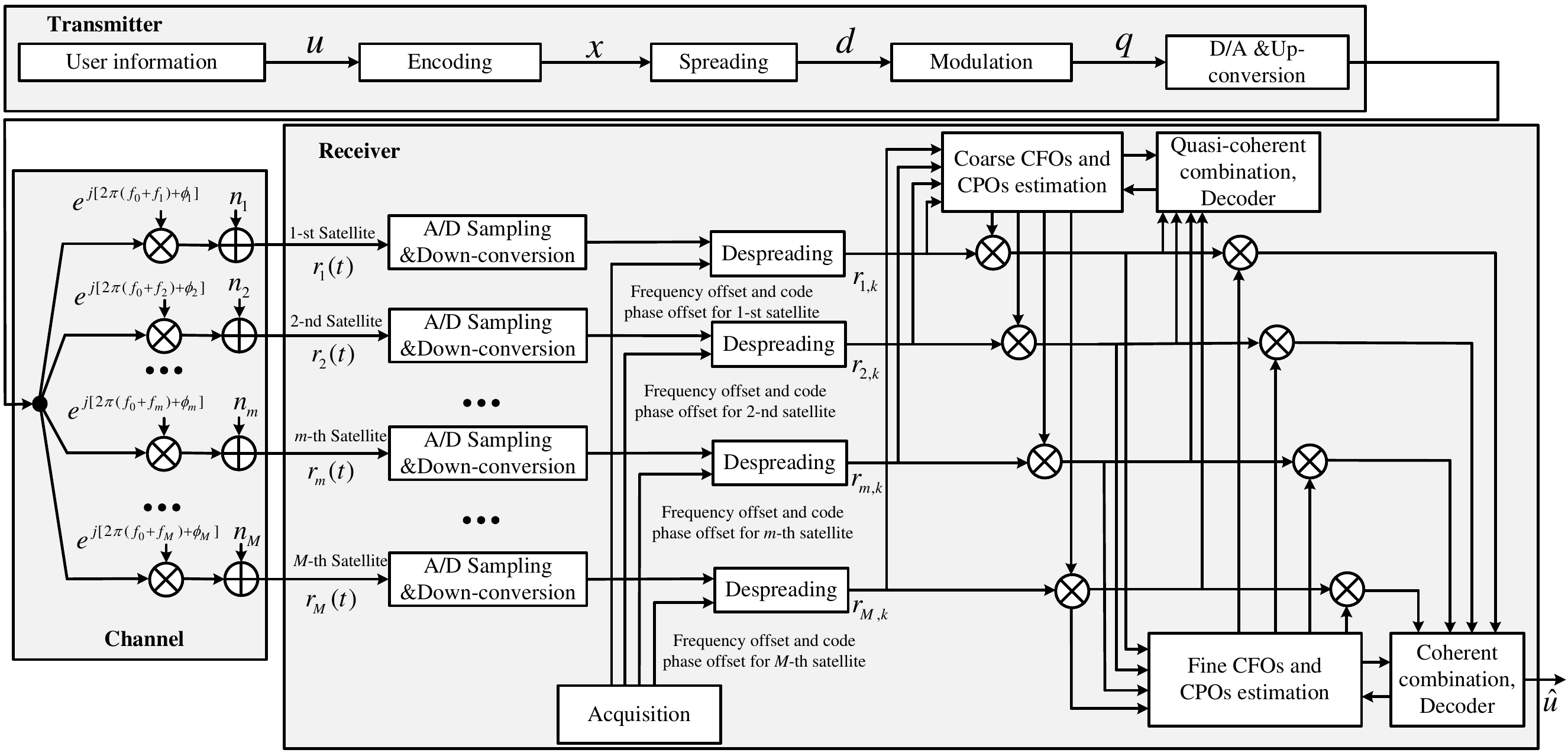}
		\caption{System architecture of uplink CSC.}
		\label{transceiver}
	\end{figure*}
	%% 图片字体有问题
	
	Direct Sequence Spread Spectrum (DSSS) is widely employed in secure satellite communication systems as it provides inherent resistance against interception, eavesdropping, and interference. 
	Among the various modulation schemes, the Binary Phase Shift Keying-Direct Sequence Spread Spectrum  (BPSK-DSSS) modulation scheme has garnered considerable attention owing to its superior performance in noisy environments. 
		To further enhance the reliability of short-burst satellite communication systems operating in noisy channels, Polar codes can be employed. 
		Polar codes belong to a class of error-correcting codes that can approach the Shannon capacity limit even with relatively short code lengths, while maintaining low complexity for both encoding and decoding processes \cite{Arikan2009Channel,Tal2015List}.  
		These codes are particularly suitable for high-reliability communications, especially in scenarios involving short-burst-frame transmission \cite{Jing2017Research}.
		The transceiver system considered in this paper is depicted in Fig.\ref{transceiver}.  
	%添加模型图
	%Taking advantage of the inherent properties of polar codes, such as their systematic structure and efficient error correction capability, the coding gain can be effectively leveraged to improve the accuracy and reliability of carrier synchronization. 
	%%改为DSSS-BPSK
	
	At the transmitter, the short frame messages to be transmitted are in the form of bit sequence denoted as $\boldsymbol{u}=(u_0,u_1,\cdots,u_{N-1})^{\mathrm{T}}$,  where $N$ is the number of the information bits. The information sequence is encoded using a binary coder with rate $R$, which generates the coded sequence $\boldsymbol{x}=(x_0,x_1,\cdots,x_{K-1})^{\mathrm{T}}$. Wherein $K=N/R$ and $R<1$. Then, the encoded frame is spread as $\boldsymbol{d}$ with the periodical spreading sequence $c(\cdot)$,  in which the time period is denoted as $T_{c}$. Based on this, the  spread sequence $\boldsymbol{d}$ is modulated into a complex symbol frame $\boldsymbol{q}=(q_0,q_1,\cdots,q_{KL-1})^{\mathrm{T}}$, where $L$ is the length of the periodical spreading sequence and the value of symbols $\boldsymbol{q}$ depends on the modulation constellation map.
		$q_{k}= \mathrm{e}^{j\pi k} \in \{1,-1\}$ represents the BPSK-modulated symbol. 	
	%	$T_{\mathrm{s}}$ is the symbol time duration, and the relation between $T_{\mathrm{c}}$ and $T_{\mathrm{s}}$ is $T_{\mathrm{s}} = K\cdot T_{\mathrm{c}}$. 
	%	Afterwards, the modulated symbols are shaped by the root raised cosine filter, whose impulse response of the filter is represented by $\bar h(\cdot)$. 
	The modulated symbol undergoes frequency up-conversion and D/A conversion. 
	The signal is then transmitted through the wireless channel and is subsequently corrupted by Additive White Gaussian Noise (AWGN).
	
	There are a total of $M$ satellites in the cooperation network, all of which are capable of receiving the signal transmitted from the same terminal. 
	The wireless channel is assumed to be a line of sight channel. 
	As a result, the received signal by a particular satellite (denoted as $m$, $m=1,2,\cdots,M$) can be written as:
	\begin{equation}\label{ReceiverDitital1}
		\begin{aligned}
			r_{m}(t) = &\sum_{k=0}^{K-1} \sum_{l=0}^{L-1}A_{m}x_{k}c\left(t-l T_{\mathrm{c}}-kT_{\mathrm{s}}\right) \\
			&\times \mathrm{e}^{j\left(2 \pi(f_{0}+f_{m})t+\phi_{m}\right)}+n_{m}(t), 
		\end{aligned}
	\end{equation}
	where $T_{\mathrm{s}}$ is the symbol time duration, and the relation between $T_{\mathrm{c}}$ and $T_{\mathrm{s}}$ is $T_{\mathrm{s}} = L\cdot T_{\mathrm{c}}$.  $f_{0}$, $f_{m}$ and $\phi_{m}$ represent the signal carrier frequency, CFO and CPO, respectively. The CFO is determined primarily by the Doppler shift, while the range of CPO varies in the interval $(-\pi,+\pi]$ due to the differences in start-up time and the drift of crystal oscillators among the collaborating satellites. The term $n_{m}(t)$ represents the complex additive Gaussian white noise at satellite $m$, with a zero mean and a variance of $\frac{N_{m}}{2}$ for both the real and imaginary components. It is assumed that the noise at each satellite is independent and uncorrelated, with equal variance. The variable $A_{m}$ denotes the received signal power from the $m$-th satellite. Throughout this paper, it is assumed that the received signal powers are equal for all satellites. Considering that the noise powers are also identical at each satellite, the SNR of the received signal is assumed to be uniform across all satellites.
	
	At the receiver, the first step in signal processing is acquisition. Typically, a low symbol rate $R_{\mathrm{s}}$ is employed during acquisition to ensure accurate signal detection while satisfying the SNR requirements. Once successful signal acquisition has been accomplished for each satellite and a preliminary estimation of the carrier frequency has been obtained. The received signal at chip level after frequency compensation becomes:
	\begin{equation}\label{ReceiverDitital2}
		\begin{aligned}
			r_{m}(t)=&\sum_{k=0}^{K-1} \sum_{l=0}^{L-1}A_{m} x_{k}c\left(t-l T_{\mathrm{c}}-k T_{\mathrm{s}}\right)\\
			&\times \mathrm{e}^{j\left(2 \pi \Delta f_{m}t +\phi_{m}\right)}+n_{m}(t).
		\end{aligned}
	\end{equation}
	
	To strike a balance between acquisition probability and resource consumption, the residual CFO after compensation based on the acquisition operation is $\Delta f_{m} \in \left(-\frac{R_{\mathrm{s}}}{2I},+\frac{R_\mathrm{s}}{2I}\right]$, where $I$ represents the number of FFT points utilized in the algorithm mentioned in \cite{IntroTang}. Consequently, the remaining NFO, obtained by multiplying the CFO with the symbol period $T_{\mathrm{s}}$ is reduced to the range of $\left(-\frac{1}{2I},+\frac{1}{2I}\right]$. The primary objective of this paper is to evaluate the performance of CFO and CPO estimation. To focus on these specific parameters, it is assumed that perfect timing recovery during despreading has been achieved and the signal amplitude has been normalized. Therefore, $r_{m}(t)$ is considered as BPSK modulation. After sampling, the $k$-th baseband complex symbol can be written as: 
			%%%% %%%做归一化所以就是等增益合并了，不具有普适性
			\begin{equation}\label{ReceiverDitital4}
				\begin{aligned}
					r_{m,k}= s_{k}\mathrm{e}^{j(2 \pi k \Delta f_{m} T_{\mathrm{s}}+\phi_{m})}+n_{m}(k).
				\end{aligned}
			\end{equation}
			where $s_{k}$ is the equivalent BPSK transmitted $k$-th symbol after despreading. 	
	%The following step is frame synchronization, which ensures that the receiver correctly identify the start of each frame of data transmitted by the sender. The length of the frame header is designated as $K$, and after implementing FFT-based frame synchronization and frequency compensation, the remaining Normalized Frequency Offset (NFO) $\Delta f_{m}T_{\mathrm{s}}$ is reduced to a range of $\mathrm{NFO}_{m} \in \left(-\frac{1}{2I},+\frac{1}{2I}\right]$.
	%% 不提帧同步，归一化频偏在上面定义	
	The $k$-th received signal symbol vector $\boldsymbol{r}_{k}$ of all the $M$ satellites can be described as,
	\begin{equation}\label{ReceiverDitital5}
		\begin{aligned}
			\boldsymbol{r}_{k}= [r_{1,k},r_{2,k},\cdots,r_{M,k}]^{\mathrm{T}}.
		\end{aligned}
	\end{equation}
	
	Thus, the received symbol vector of all the $M$ satellites is:
	\begin{equation}\label{ReceiverDitital6}
		\begin{aligned}
			\mathbf{r}= [\boldsymbol{r}_{0}^{\mathrm{T}},\boldsymbol{r}_{1}^{\mathrm{\mathrm{T}}},\cdots,\boldsymbol{r}_{K-1}^{\mathrm{T}}]^{\mathrm{T}}.
		\end{aligned}
	\end{equation}
	% Assuming that perfect timing recovery have been achieved at each satellite, the complex samples at the output of the matched filter can be expressed as:
	% \begin{equation}\label{ReceiverDitital4}
		% 	\begin{aligned}
			% 		r_{m,k}= s_{k}\mathrm{e}^{j(2 \pi k \Delta f_{m} T_{\mathrm{s}}+\phi_{m})}+n_{m}(k).
			% 	\end{aligned}
		% \end{equation}
	%%系统图的后半部分没有描述

		\section{Problem Formulation}
		Accurately estimating the CFOs and CPOs of all the $M$ satellites is essential for achieving coherent  combination in CSC. These parameters can be denoted as:
		\begin{align}\label{ReceiverDitital6}
			\boldsymbol{\theta} &= [\mathrm{CFO}_{1},\cdots,\mathrm{CFO}_{m},\mathrm{CFO}_{M}, \phi_{1},\phi_{2},\cdots,\phi_{M}]^{\mathrm{T}} \notag \\
			&= [\boldsymbol{f}, \boldsymbol{\phi}]^{\mathrm{T}}.
		\end{align}
		
		MLE is a well-established and effective technique used to solve parameter estimation problems. This method involves computing the probability density function of the parameter being estimated. In the context of CSC, where the received signals from each satellite are encoded and modulated using identical information bit sequences, and the noise in each satellite is both independent and unrelated, the conditional probability density function of the symbol can be expressed as \eqref{pdf00}.
		\begin{figure*}[ht]
			\centering
			\begin{align}
				\label{pdf00}
				p\left(\boldsymbol{r}_{k} \mid \boldsymbol{\theta}, s_{k}\right)
				&=\left(\pi^{M} \prod_{m=1}^{M} N_{m}\right)^{-1}
				\exp \left \{ -\sum_{m=1}^{M} \frac{|r_{m, k}- s_{k}\mathrm{e}^{j(2 \pi k \Delta f_{m} T_{\mathrm{s}}+\phi_{m})}|^2}{N_{m}} \right \} \notag \\
				&=\left(\pi^{M} \prod_{m=1}^{M} N_{m}\right)^{-1}
				\exp \left \{-\sum_{m=1}^{M} \frac{(|r_{m, k}|^2 - 2 \Re\left[ s_{k}^{*}\mathrm{e}^{-j(2 \pi k \Delta f_{m} T_{\mathrm{s}}+\phi_{m})} r_{m, k}\right ]  + |s_{k}|^2)}{N_{m}} \right \} 
			\end{align}
			\rule{18cm}{0.01cm}
		\end{figure*}
		
		By removing the constant component, which is irrelevant to the parameter estimation, \eqref{pdf00} can be simplified to:
		\begin{equation}
			\label{pdf1}
			\begin{aligned}
				p\left(\boldsymbol{r}_{k} \mid \boldsymbol{\theta}, s_{k}\right) 
				&=\exp \left \{ \sum_{m=1}^{M} \frac{2 \Re\left [s_{k}^{*}\mathrm{e}^{-j(2 \pi k \Delta f_{m} T_{\mathrm{s}}+\phi_{m})} r_{m, k}\right ] }{N_{m}} \right \}.
			\end{aligned}
		\end{equation} 
		
		The data symbol $s_{k}$ is an independent and identically distributed discrete random variable, and its probability function can be represented as $p_{s}(s_{k})$. 
		By applying the law of total probability and \eqref{pdf1}, the probability density function of the $k$-th symbol vector $\boldsymbol{r}_{k}$ can be derived as:
	    \begin{align}\label{pdf2}
			p\left(\boldsymbol{r}_{k} \mid \boldsymbol{\theta}\right)&=\mathbb{E}_{s_{k}}\left[p\left(\boldsymbol{r}_{k} \mid \boldsymbol{\theta}, s_{k}\right)\right] \notag \\
			&=\sum_{s_{k} \in \mathfrak{M}} p_{k}(\mathfrak{m}) p\left(\boldsymbol{r}_{k} \mid \boldsymbol{\theta}, s_{k}\right). 
		\end{align}
		$\mathfrak{M}$ represents the modulation order.  In terms of BPSK-modulated signal, $\mathfrak{M} = 2$ and it has
		\begin{equation}\label{pdf3}
			p_{k}(\mathfrak{m}) = \frac{1}{2}, \mathfrak{m} = \left\{0,1\right\}.
	\end{equation}
	
	Combining \eqref{pdf1} to \eqref{pdf3}, the probability density function of the $k$-th data vector $\boldsymbol{r}_{k}$ is expressed as:
	\begin{equation}\label{pdf4}
		\begin{aligned}
			p\left(\boldsymbol{r}_{k} \mid \boldsymbol{\theta}\right)=\cosh \left \{ \sum_{m=1}^{M} \frac{2 \Re\left [ \mathrm{e}^{-j(2 \pi k \Delta f_{m} T_{\mathrm{s}}+\phi_{m})} r_{m, k}\right ] }{N_{m}}\right\}.
		\end{aligned}
	\end{equation}
	
	Due to the independence of $s_{k}$, the probability density function of the data vector of all the $M$ satellites is written as:	
	\begin{align}\label{pdf5}
		p(\mathbf{r}\mid\boldsymbol{\theta})&=\prod_{k=0}^{K-1} p\left(r_{k} \mid \boldsymbol{\theta}\right) \notag\\
		&=\prod_{k=0}^{K-1} \left\{ \cosh \left \{\sum_{m=1}^{M} \frac{2 \Re\left [ \mathrm{e}^{-j(2 \pi k \Delta f_{m} T_{s}+\phi_{m})} r_{m, k}\right ] }{N_{m}}\right \}\right\}.
	\end{align}
	
	Accordingly, the MLE of $\boldsymbol{\theta} = [\boldsymbol{f}, \boldsymbol{\phi}]^{\mathrm{T}}$ can be expressed as: 
	\begin{equation}\label{pdf6}
		\begin{aligned}
			\hat{\boldsymbol{\theta}} = \arg \max_{\boldsymbol{\theta \in \mathcal{F}}} [\ln p(\boldsymbol{r} \mid \boldsymbol{\theta})], 
		\end{aligned}
	\end{equation}
	where $\mathcal{F}$ is the range of the parameter $\boldsymbol{\theta}$.
	%	where $\mathcal{F} = \left \{  \mathrm{NFO}_{1},\cdots,\mathrm{NFO}_{m},\mathrm{NFO}_{M},\phi_{1},\phi_{m},\cdots,  \phi_{M}\right \}$, $\mathrm{NFO}_{m} \in \left(-\frac{1}{2I},+\frac{1}{2I}\right], \phi_{m} \in (-\pi,+\pi]$.
	
		It is observed that obtaining the MLE value from the log likelihood function is a challenging and nonlinear optimization problem, primarily due to non-linearity, a large search space, and the presence of multiple local minima. These factors make direct solutions difficult. In response, researchers have explored iterative approaches to address this issue, such as Sumple \cite{Rogstad2005TheSA,Rogstad2007AligningAA}, and Simple \cite{Rogstad2005TheSA}.
  %, and Particle Swarm Optimization Frequency and Phase Estimation(PSOFPE) \cite{Ke2011Multi-parameter}. 
  %中文文献
  The SIMPLE algorithm achieves array coherence by employing a simple pair-wise correlation of antenna signals, while the SUMPLE algorithm performs a summation operation on multiple antenna signals prior to correlation. Both algorithms assume perfect compensation of carrier frequency offset and aim to estimate phase deviations through iterative cross-correlation of multi-channel received signals, maximizing combined SNR. Continuous signal updating is required during iteration, making these algorithms more suitable for continuous communication scenarios rather than short-burst communication. 
  %To address this limitation, the PSOFPE algorithm has been proposed as an alternative. This algorithm randomly assigns values to carrier frequency and phase offsets, and maximizes equation \eqref{pdf6} within a limited data length. However, it falls in the category of NDA method, which results in poor estimation accuracy \cite{Du2021Optimum}. 
  On the other hand, the CA method offers comparable estimation accuracy to DA methods, making it more desirable for limited data transmission scenarios. Nonetheless, it is susceptible to residual CFO and CPO, potentially leading to encoding and decoding errors as well as communication failure. Overall, for the considered communication system, there currently is not a suitable solving method available. 
	%这里我觉得要从这几个方面来说
	%(1) 非线性直接求解非常困难  simple等迭代 但是不适合短帧猝发 并且基于非数据辅助 估计性能较低 编译码性能好 
	
	%(2) 极低信噪比 比常规的单星信噪比还要低 
	
	%(3) 参数估计范围大 尤其是载波相位 但是受参数估计范围影响
	
	\section{Proposed CFO and CPO Estimation Algorithm}
\begin{figure*}[]
		\centering
		\includegraphics[width=1\textwidth]{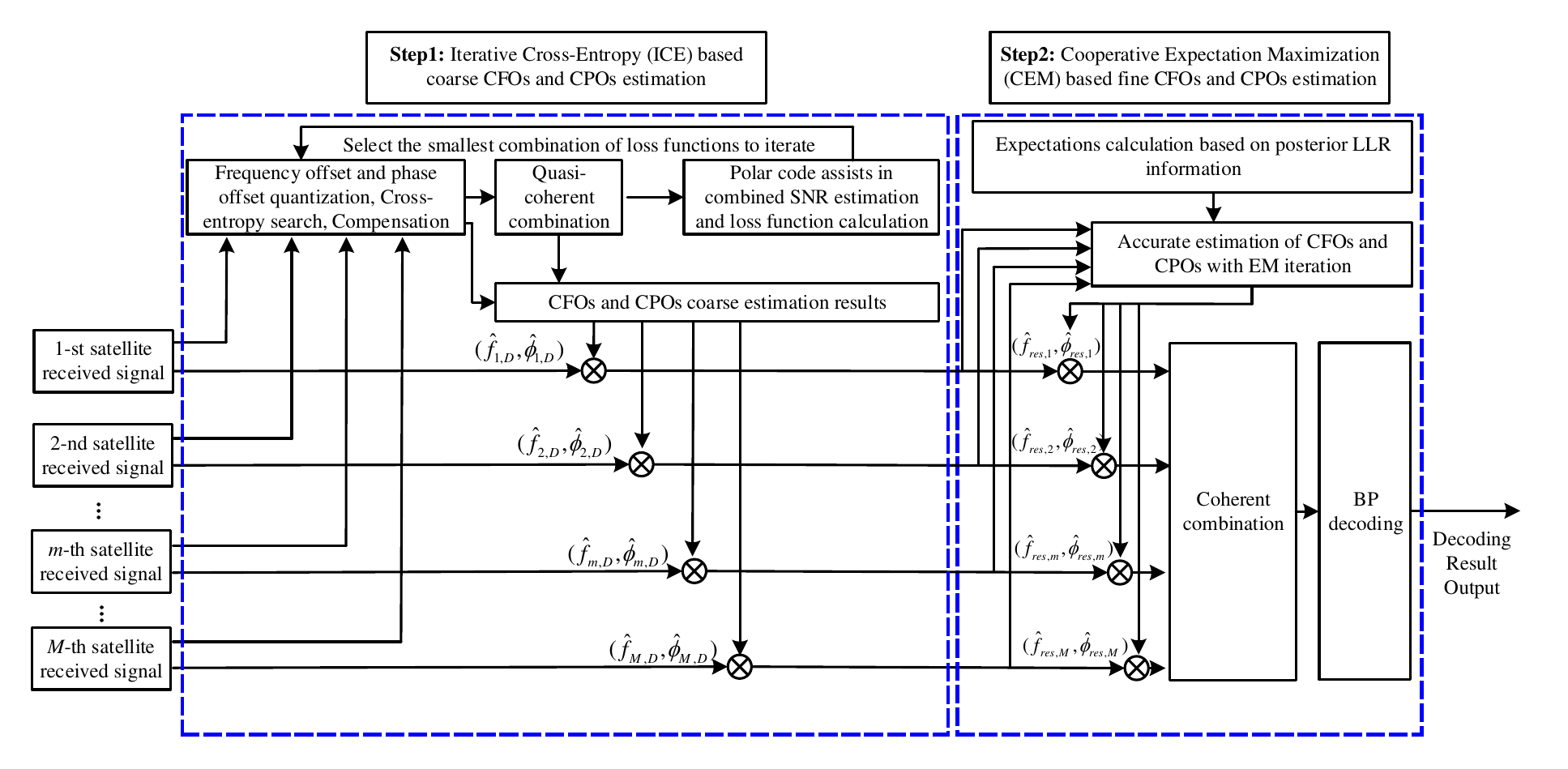}
		\caption{The schematic diagram of the estimator.}\label{CFOCPO1}
	\end{figure*}

	To ensure signal coherent combination in a CSC system under low SNR conditions, we propose an iterative CA estimation algorithm for joint CFOs and CPOs estimation. The algorithm consists of two parts: Iterative estimation based on Cross Entropy  (ICE) and Cooperative Expectation Maximization (CEM). The ICE is responsible for coarse search of CFOs and CPOs, while the CEM is responsible for CFOs and CPOs fine estimation. The overall structure of these two parts is depicted in Fig.\ref{CFOCPO1}. The underlying principles of these two parts will be elaborated upon in the following subsections.
	
		\begin{figure*}
		\centering
		\includegraphics[width=0.75\textwidth]{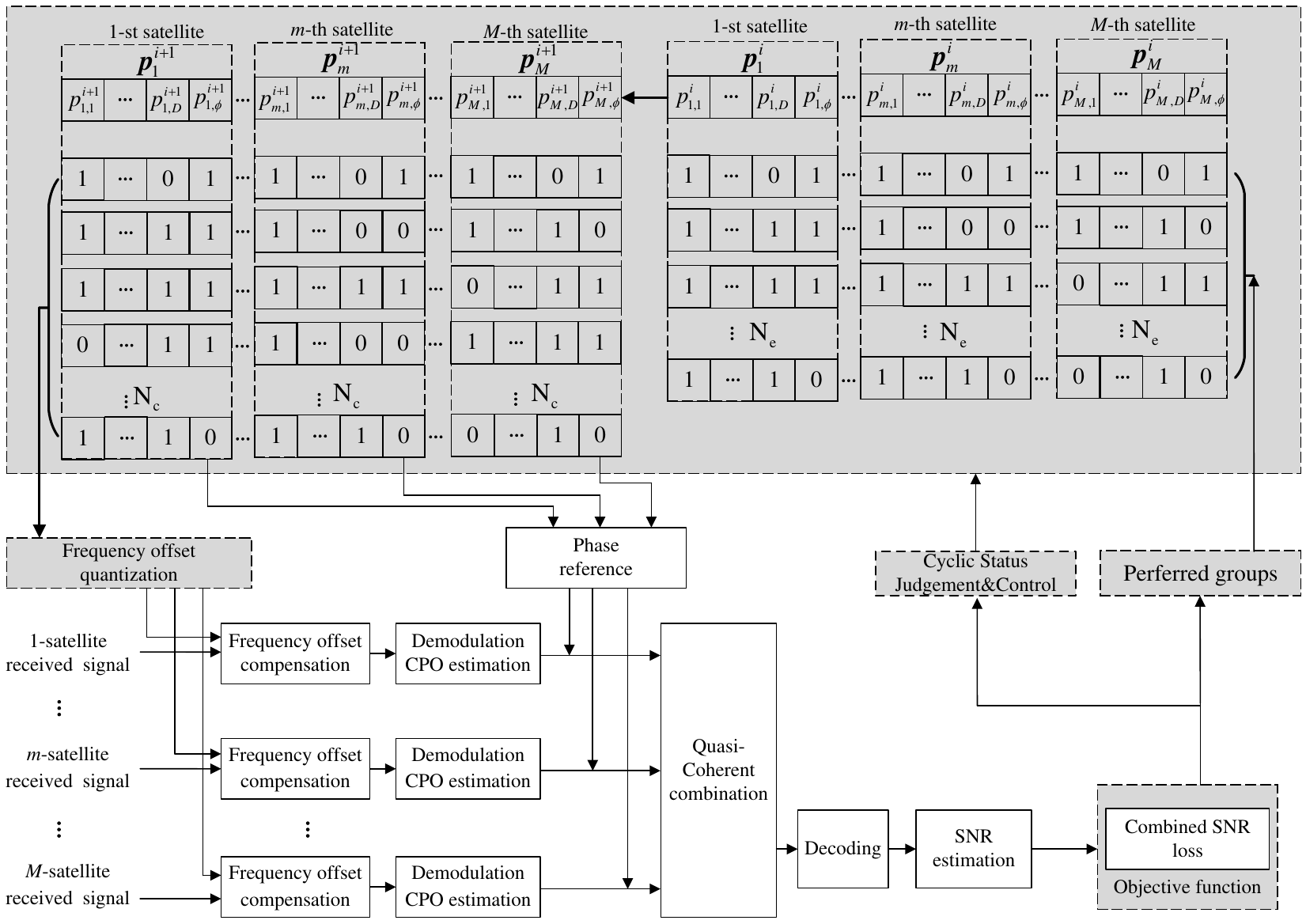}
		\caption{The schematic diagram of the ICE algorithm.}\label{PolarCrossEntroy}
	\end{figure*}
	%图片修改 字体，添加Step1和step2

	\subsection{Iterative Cross-Entropy based Coarse CFOs and CPOs Estimation}
	The initialization step of our proposed algorithm involves the utilization of a cross-entropy-based iteration method, which is commonly employed in parameter estimation tasks. This method measures the disparity between the true probability distribution and the predicted probability distribution of a given model, making it an efficient parallel search algorithm. It offers several advantages, including simplicity, ease of implementation, fewer parameters, and expandability, making it suitable for our purposes. In the context of CFO and CPO estimation, our algorithm aims to mitigate the uncertainties associated with their ranges. The CFO range falls within the interval of $\left(-\frac{1}{2I}, +\frac{1}{2I}\right]$, while the CPO range falls within $(-\pi,\pi]$. By applying the cross-entropy method, we can effectively conduct a parallel search to estimate these parameters.	Our proposed ICE algorithm is depicted in Fig.\ref{PolarCrossEntroy}. To handle CFO estimation, we employ quantization with $D$ bits, which allows us to compensate  the received signal at each satellite using the quantized frequency offset. Subsequently, demodulation is performed to calculate the phase offset. However, the square demodulation of a BPSK-modulated signal introduces a phase ambiguity, which we resolve by introducing an additional bit. As a result, the cross-entropy iteration process involves the participation of the received signal at each satellite using $D+1$ bits. During the iteration process, the objective function that drives the iteration and defines the convergence condition is the combined SNR loss. We select the optimal CFOs and CPOs when the objective function reaches its minimum. Accurate SNR estimation is crucial for optimization; hence, we incorporate a Polar code-aided estimation algorithm to enhance the precision of SNR estimation. By employing this approach, we can successfully complete the joint search for CFOs and CPOs of multiple satellites within a few iterations.

	\subsubsection{Objective Function}
	
	Let $\boldsymbol{b}_{1\times M(D+1)} = [\boldsymbol{b}_{1},\boldsymbol{b}_{2},\cdots,\boldsymbol{b}_ {m},\boldsymbol{b}_{M}]$ represent the quantized binary set of CFOs and CPOs of the received signal at all $M$ satellites. Here $\boldsymbol{b}_{m} = [b_{m,1},b_{m ,2},\cdots,b_{m,d},b_{m,D}, b_{m,\phi}]_{1\times(D+1)}$, with $b_{m,d} \in [0,1]$, represents the $D+1$ bits quantized binary set of the $m$-th satellite, $d \in [1,D]$. Besides, $b_{m,\phi} \in [0, 1]$ represents the quantized value of the CPO of the received signal at the $m$-th satellite. For simplicity, let $l =1,2,\cdots,M(D+1)$ represent the index of the quantized bit. After  combining the signals, the $k$-th symbol is recorded as $r_{\mathrm{com},k}$, which can be expressed in detail as:
	\begin{equation}\label{ReceiverDitital8}
		\begin{aligned}
			r_{\mathrm{com},k} = \sum^{M}_{m=1}r_{m,k}\mathrm{e}^{-j[2 \pi f_{m,d} k T_{\mathrm{s}}+ (2b_{m,\phi}-1)\pi + \phi_{m,d}]},
		\end{aligned}
	\end{equation}
	where $f_{m,d}$ is the CFO estimation of the received signal at the $m$-th satellite. In this case, $f_{m,d}$ is written as:
	\begin{equation}\label{Dopplerm}
		\begin{aligned}
			f_{m,d} = \left(\sum\limits_{d=1}^{D}2^{b_{m,d}}-2^{D}\right)\frac{R_{\mathrm{s}}}{I2^{D}}.
		\end{aligned}
	\end{equation}
	
	Additionally, $\phi_{m,d}$ in \eqref{ReceiverDitital8} is the compensated result of the received signal at the $m$-th satellite, which is expressed by:
	\begin{equation}\label{CPOEst}
		\begin{aligned}
			\phi_{m,d}&=\frac{1}{2} \arg \left(\sum_{k=0}^{K-1} r_{m,k}^{2} \mathrm{e}^{-j 2 \pi 2 f_{m,d} kT_{\mathrm{s}}}\right).
		\end{aligned}
	\end{equation}
	In order to eliminate the phase ambiguity and achieve the coherent combination, each satellite introduces $1$ bit to indicate selection between $\phi_{m,d}$ and $\phi_{m,d} +\pi$.

		Assuming that the SNR of each satellite as $\mathrm{SNR_{Single}}$, the theoretical SNR after coherent combination is given by $\mathrm{SNR_{Single}} + 10\log_{10}(M)$. In this context, the measure of combined SNR loss can be expressed as a function involving the estimated combined SNR $\hat{\gamma}$:
		\begin{equation}\label{SNRLoss}
			\begin{aligned}
				\text{SNRLoss} = \mathrm{SNR_{Single}} + 10\log_{10}(M) - \hat{\gamma},
			\end{aligned}
		\end{equation}
		
		The calculation of the estimated combined SNR of a BPSK-modulated signal using the CA method can be expressed as:
		\begin{equation}\label{SNREst1}
			\begin{aligned}
				\hat{\gamma}=\frac{\left(K-\frac{3}{2}\right)\left(\sum\limits_{k=0}^{K-1} \Re\left\{r_{\mathrm{com},k} \zeta_{k}^{*}\right\}\right)^{2}}{K\left \{K\left(\sum\limits_{k=0}^{K-1}\left|r_{\mathrm{com},k}\right|^{2}\right)-\left(\sum\limits_{k=0}^{K-1} \Re\left\{r_{\mathrm{com},k} \zeta_{k}^{*}\right\}\right)^{2}\right \}},
			\end{aligned}
		\end{equation}
		Where $\zeta_{k}$ is the posterior expectation of the $k$-th combined signal, and the symbol $*$ denotes the conjugation operation.
		
		The objective function aims to minimize the SNR loss by finding the optimal binary set $\boldsymbol{b}$, namely:
		\begin{equation}\label{Crossentrpy1}
			\begin{aligned}
				\hat{\boldsymbol{b}} = \arg \min_{\boldsymbol{b}\in \mathcal{B}}(\text{SNRLoss} < T_{\mathrm{H}}),
			\end{aligned}
		\end{equation}
		Here, the notation $T_{\mathrm{H}}$ represents the threshold value. Additionally, $\mathcal{B}$ denotes the complete set of quantized CFOs and CPOs of all $M$ satellites. 
		
		\subsubsection{Algorithm Design} 
		The parameter $\zeta_{k}$ in \eqref{SNREst1} can be calculated by the posterior Log-Likelihood Ratio (LLR) from the output of the decoder. The posterior probability of the encoded data $x_{k}$ is expressed as:
			\begin{equation}\label{LLR1}
				\begin{aligned}
					L_{\mathrm{pos}}(x_{k}) =\ln \frac{\text{Pr}\left\{{x_{k} = 0|r_{\mathrm{com},k}}\right\}}{\text{Pr}\left\{{x_{k} = 1|r_{\mathrm{com},k}}\right\}},
				\end{aligned}
			\end{equation}
			without loss of generality, for BPSK-modulated signal, there is:
			\begin{equation}\label{LLR2}
				\begin{aligned}
					\text{Pr}\left\{{s_{k} = \mathrm{e}^{j\cdot 0\cdot\pi}|r_{\mathrm{com},k}}\right\} = \text{Pr}\left\{{{x}_{k}= 0 |r_{\mathrm{com},k}}\right\},
				\end{aligned}
			\end{equation}
			\begin{equation}\label{LLR3}
				\begin{aligned}
					\text{Pr}\left\{{s_{k} = 0|r_{\mathrm{com},k}}\right\} + \text{Pr}\left\{{{x}_{k} = 1|r_{\mathrm{com},k}}\right\} = 1.
				\end{aligned}
			\end{equation}
			By combining equations \eqref{LLR1}-\eqref{LLR3}, we can derive the conditional probability of $s_{k} = \mathrm{e}^{j\cdot 0 \cdot\pi}$ given $r_{\mathrm{com},k}$ as:
			\begin{equation}\label{LLR4}
				\begin{aligned}
					\text{Pr}\left\{{s_{k} = \mathrm{e}^{j\cdot 0 \cdot\pi}|r_{\mathrm{com},k}}\right\} = \frac{\mathrm{e}^{L_{\mathrm{pos}}(x_{k})}}{\mathrm{e}^{L_{\mathrm{pos}}(x_{k})} + 1}.
				\end{aligned}
			\end{equation}
			Similarly, the conditional probability of $s_{k} = \mathrm{e}^{j\cdot 1 \cdot\pi}$ given $r_{\mathrm{com},k}$ can be expressed as: 
			\begin{equation}\label{LLR5}
				\begin{aligned}
					\text{Pr}\left\{{s_{k} = \mathrm{e}^{j\cdot 1 \cdot\pi}|r_{\mathrm{com},k}}\right\} = \frac{1}{\mathrm{e}^{L_{\mathrm{pos}}(x_{k})} + 1}.
				\end{aligned}
			\end{equation}
			Base on this, by combining \eqref{LLR4} and \eqref{LLR5}, the expectation of $\zeta_{k}$ can be described as:
			\begin{align}\label{LLR6}
				\zeta_{k} =& \mathrm{e}^{j\cdot 0 \cdot\pi} \cdot \text{Pr}\left\{{s_{k} = \mathrm{e}^{j\cdot 0 \cdot\pi}|r_{\mathrm{com},k}}\right\} \notag \\
				&+ \mathrm{e}^{j\cdot 1 \cdot\pi} \cdot \text{Pr}\left\{{s_{k} = \mathrm{e}^{j\cdot 1 \cdot\pi}|r_{\mathrm{com},k}}\right\} \notag \\
				=& 1 \cdot \frac{\mathrm{e}^{L_{\mathrm{pos}}(x_{k})}}{\mathrm{e}^{L_{\mathrm{pos}}(x_{k})} + 1} + (-1)\cdot  \frac{1}{\mathrm{e}^{L_{\mathrm{pos}}(x_{k})} + 1}\notag \\
				=&  \frac{\mathrm{e}^{L_{\mathrm{pos}}(x_{k})}-1}{\mathrm{e}^{L_{\mathrm{pos}}(x_{k})} + 1} \notag \\ =&  \tanh\left(\frac{\mathrm{e}^{L_{\mathrm{pos}}(x_{k})}}{2}\right),
			\end{align}
			where $\tanh$ represents the hyperolic tangent function, namely, $\tanh(x) = \frac{\mathrm{e}^{x}-\mathrm{e}^{-x}}{\mathrm{e}^{ x}+\mathrm{e}^{-x}}$. The computation of the hyperbolic tangent function involves a relatively high complexity. To mitigate this, a common approach is to approximate it using a linear piecewise function. This approximation can be mathematically represented as:
			\begin{equation}\label{Lpos}		
				\begin{aligned}
					\zeta_{k}=	\begin{cases}1, L_{\mathrm{pos}}(x_{k})>T_{h}\\
						\alpha L_{\mathrm{pos}}(x_{k}), -T_{h}<L_{\mathrm{pos}}(x_{k})\leq T_{h}\\
						-1,  L_{\mathrm{pos}}(x_{k})\leq -T_{h},\\
					\end{cases}
				\end{aligned}
			\end{equation}
			where $L_{\mathrm{pos}}(x_{k})$ indicates the posterior LLR of the $k$-th symbol $r_{\mathrm{com},k}$, $\alpha$ represents the linear factor, and $T_{h}$ denotes the segmentation point. To achieve a more accurate approximation of the hyperbolic tangent function, the value of $\alpha$ is set to $1/3$ and $T_{h}$ is set to $3$, as suggested in the literature  \cite{chen2019intelligent}. 
		
		In Fig.~\ref{PolarCrossEntroy}, the probability vector for the $i$-th iteration process is defined as $\boldsymbol{p}^{i}_{1\times M(D+1)} = [\boldsymbol{p}_{1}^{i},\cdots,\boldsymbol{p}_{m}^{i},\cdots,\boldsymbol{p}_{M}^{i}]$. Here $\boldsymbol{p}_{m}^{i}$ represents the probability of CFO and CPO $D+1$ bits quantization of the $m$-th satellite. To simplify the notation, we denote $p^{i}_{l}$ an element in $\boldsymbol{p}^{i}$, where $l \in 1,\cdots, M(D+1)$. Subsequently, the probability vector $\boldsymbol{p}^{i}_{1\times M(D+1)}$ is utilized to randomly generate a set of candidate observation matrices $\left \{ \mathbf{B}\right \}_{n_{\mathrm{c}}=1}^{N_{\mathrm{c}}}$. Here $N_{\mathrm{c}}$ denotes the number of vectors. These $N_{\mathrm{c}}$ groups of candidate vectors are then combined with multiple satellites using equations \eqref{ReceiverDitital8}-\eqref{CPOEst}, and the objective function is calculated to obtain the combined SNR loss vector $\text{SNRLossVec}_{1 \times N_{\mathrm{c}}}$. Based on this, the combined SNR loss vector is arranged in ascending order as $\eta_\text{seq,1}^i \leq \eta_\text{seq,2}^i \leq,\cdots,\leq \eta_{\text{seq}, N_\text{c}}^i$. The $N_{\mathrm{e}}$ vectors with the smallest combined SNR loss are then selected and recorded as $\left \{ \mathbf{B}\right \}_{n_{\mathrm{c}} =d_{\mathrm{e},1}}^{ n_{\mathrm{c}} =d_{\mathrm{e},N_{\mathrm{e}}}}$, where $d_{\mathrm{e},n_{\mathrm{e}}}$ represents the index of the selected vectors. Subsequently, the probability generation vector is updated based on these indices.	 
		\begin{equation}\label{jiaochashangdiedai}		
			\begin{aligned}
				\boldsymbol{p}^{i+1} = (1-\bar{w})\boldsymbol{p}^{i} + \frac{\bar{w}}{N_{\mathrm{e}}}(\boldsymbol{b}_{d_{\mathrm{e},1}}+\boldsymbol{b}_{d_{\mathrm{e},2}},\cdots,\boldsymbol{b}_{d_{\mathrm{e},N_{\mathrm{e}}}}),
			\end{aligned}
		\end{equation}
		By introducing a smoothing parameter $\bar{w}$, the iterative process is carried out for a total of $N_{\mathrm{iter}}$ iterations to obtain the optimal solution $\mathbf{b}^{*}$. Based on this, the estimated matrix$\hat{\boldsymbol{\Phi}}_{D} = [\hat{\boldsymbol {f}}_{D}, \hat{\boldsymbol{\phi}}_{D}]^{T}$ can be obtained. Here $\hat{\boldsymbol{f}}_{D} = [\hat{f }_{1,d},\hat{f}_{2,d},\cdots,\hat{f}_{m,d},\hat{f}_{M,d}]$ and \\ $ \hat{\boldsymbol{\phi}}_{D} = [\hat{\phi}_{1,d},\hat{\phi}_{2,d},\cdots,\hat{\phi }_{m,d},\hat{\phi}_{M,d}]$ represent the estimated CFO and CPO parameters for each satellite. The algorithmic procedure described above is summarized in Algorithm 1.
	
	\begin{algorithm}
		\SetAlgoLined
		\KwIn{$N_{\mathrm{c}}$, $N_{\mathrm{e}}$, $N_{\mathrm{iter}}$, received signal of all $M$ satellites $\mathbf{r} = [\boldsymbol{r}_{1},\boldsymbol{r}_{2},\cdots,\boldsymbol{r}_{M}]^{\mathrm{T}}$}
		\KwOut{$\hat{\boldsymbol{b}}$, $\hat{\boldsymbol{\Phi}}_{D}$}
		
		Initialization: Set the initial probability for the CFOs and CPOs quantized bits of each satellite as $\boldsymbol{p}^{i} = 0.5*\mathbf{1}_{1\times M(D+1)}$, where $i$ = 0;\\
		\While{$i\leq N_{\mathrm{iter}}-1$}{
			(1) Generate observation matrices $\left \{ \mathbf{B}\right \}_{n_{\mathrm{c}}=1}^{N_{\mathrm{c}}}$ to obtain the quantized CFOs and CPOs of received signal symbol at each satellite according to \eqref{Dopplerm} and \eqref{CPOEst};\\
			(2) Calculate the combined SNR loss denoted as $\text{SNRLossVec}_{1  \times N_{\mathrm{c}}}$, and mark its elements as $\eta_1^i,\eta_2^i,\ldots,\eta_{N_\text{c}}^i$;\\
			(3) Sort the elements in ascending order and update it as $\eta_\text{seq,1}^i \leq \eta_\text{seq,2}^i \leq,\cdots,\leq \eta_{\text{seq},N_\text{c}}^i$;\\
			(4) Select the top $N_{\mathrm{e}}$ with the smallest combined SNR loss and calculate the iteration probability of the next iteration by $\boldsymbol{p}^{i+1} =(1-\bar{w})\boldsymbol{p}^{i} + \frac{\bar{w}}{N_{\mathrm{e}}}(\boldsymbol{b}_{d_{\mathrm{e},1}}+\boldsymbol{b}_{d_{\mathrm{e},2}},\ldots,\boldsymbol{b}_{d_{\mathrm{e},N_{\mathrm{e}}}})$;\\
			(5) $i = i + 1$;\\
		}
		return:$\hat{\boldsymbol{b}}$, $\hat{\boldsymbol{\Phi}}_{D}$
		\caption{ICE algorithm}
		\label{euclid}
	\end{algorithm}
	
	\subsubsection{Performance Simulation}
	According to the aforementioned algorithm design, the selection of parameters such as the frequency offset quantization bit width $D$, the number of candidate vectors $N_ {\mathrm {c}} $, the number of elite vectors $N_ {\mathrm {e}} $, and the number of iterations $N_ {\mathrm {iter}} $ is an open question that needs to be determined. In order to analyze the impact of these parameters on the algorithm performance, we conduct simulations using the ICE algorithm for joint estimation of CFOs and CPOs.
	
\begin{figure*}[ht]
		\centering
		\subfigure [BER versus $E_{\mathrm{b}}/N_{\mathrm{0}}$ at varying normalized frequency offsets]{
			\label{SinglePolarA1}
			\includegraphics [width=0.48\textwidth]{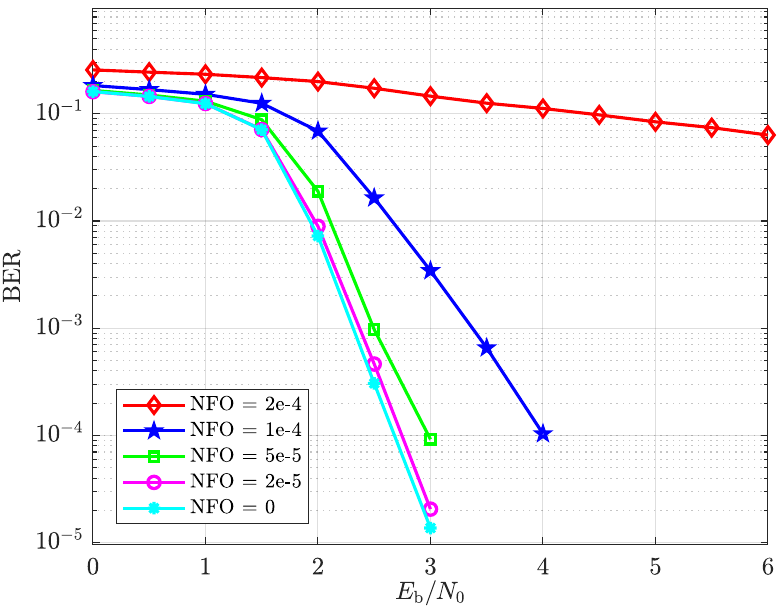}}
		\subfigure [BER versus phase offset at varying $E_{\mathrm{b}}/N_{\mathrm{0}}$]{
			\label{SinglePolarb1}
			\includegraphics [width=0.48\textwidth]{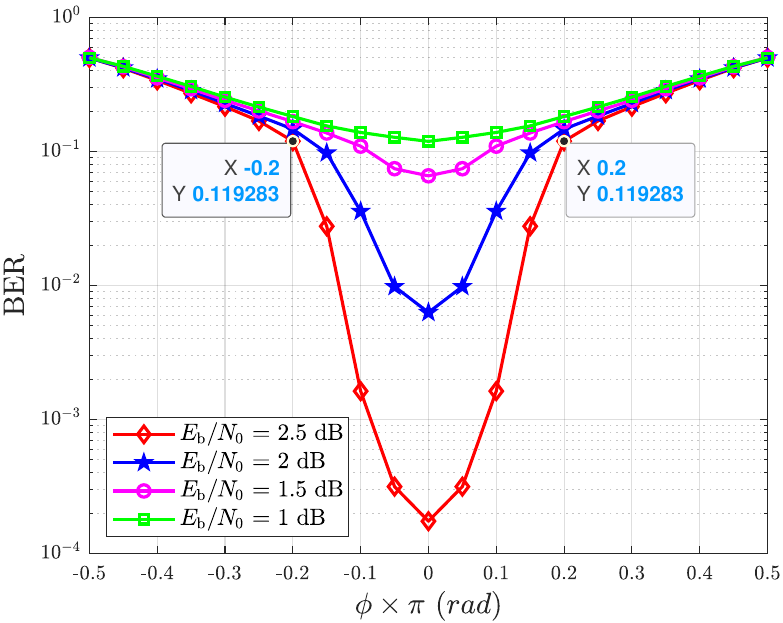}}
		\caption{The impact of CFO and CPO on the decoding performance of Polar code $\mathcal{C}$(1024,512)}
		\label{SinglePolar}
	\end{figure*}

	The choice of quantization bit width for the frequency offset requires finding a balance between system performance and complexity. To illustrate the impact of CFO and CPO on the decoding performance of Polar code $\mathcal{C}$(1024,512) using the Belief Propagation (BP) decoding algorithm \cite{PolarBP}, we perform simulations and the obtained results are shown in Fig. \ref{SinglePolar}.
	Fig. \ref{SinglePolar}(a) demonstrates that even with a small frequency offset value ($\mathrm{NFO} = 1\times 10^{-4}$), there is already a noticeable degradation of more than $1$ dB in the decoding performance (@BER = $10^{-4}$) compared to the ideal case. As the NFO further increases, the decoding performance deteriorates gradually. When NFO  $ =2\times 10^{-4}$ and $E_{\mathrm{b}}/N_{\mathrm{0}}= 1.5$ dB, the BER approaches $0.5$, indicating a significant degradation in the decoding performance and rendering the decoder ineffective.
	In Fig. \ref{SinglePolar}(b), the impact of CPO on the BER is illustrated. It can be observed that even small changes in the phase offset within the range of $[-0.2 \pi,+0.2 \pi]$ can have a considerable impact on the BER. Therefore, given the NFO value of $1\times 10^{-4}$, the quantization bit width $D$ can be determined by:
	\begin{equation}\label{DCal}
		\begin{aligned}
			D \geq \left \lfloor \log_2 \frac{1}{\mathrm{NFO} \cdot I} \right \rfloor, D\in \mathbb{N}^{+} 
		\end{aligned}
	\end{equation}
	
\begin{figure}[ht]
		\centering
		\includegraphics[width=0.48\textwidth]{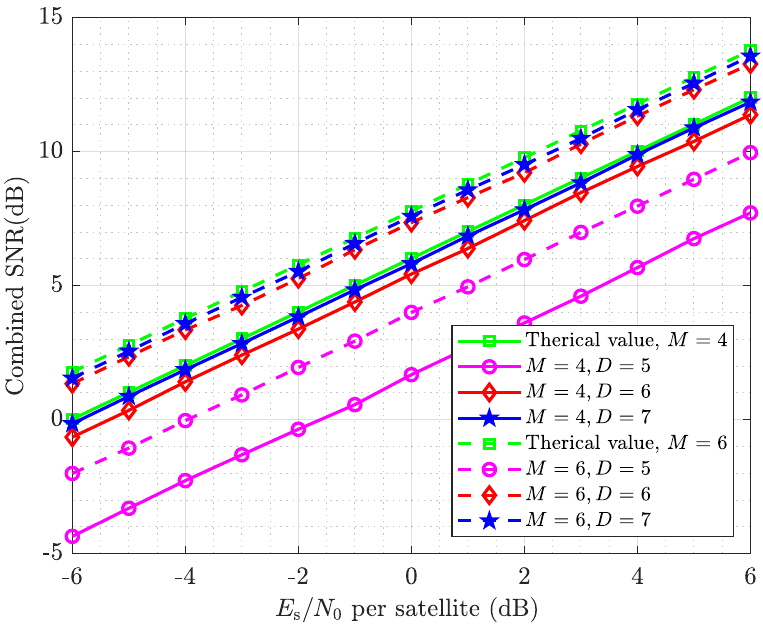}
		\caption{The relationship between combined SNR and single satellite SNR for varying frequency offset quantization bit widths in the context of multiple satellite systems.}
		\label{CombiningSNR}
	\end{figure}

\begin{figure}
		\centering
		\includegraphics[width=0.48\textwidth]{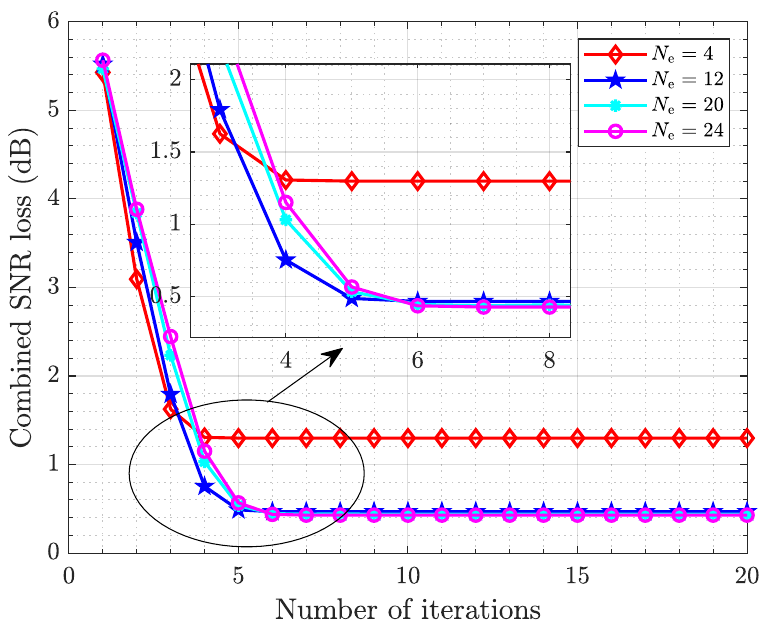}
		\caption{The relationship between $N_{\mathrm{c}}$, $N_{\mathrm{e}}$ and the iterations times}\label{Ne}
	\end{figure}

	When $R_{\mathrm{s}}=1000 $ sps and $I=64$, the range of NFO for each satellite is within the interval ($-7.8125\times10^{-3}$, $+7.8125\times10^{-3}$]. In the simulation presented in Fig.~\ref{CombiningSNR}, we consider a random combination of  $N_ {\mathrm {c}}=120$  with a candidate group size of  $N_ {\mathrm {e} }=24$. We investigate the relationship between the combined SNR of multiple satellites and the SNR of a single satellite under different frequency offset quantization bit conditions. As shown in Fig.~\ref{CombiningSNR}, when the frequency offset quantization bit width is set to $5$, the combined SNR loss for CSC with $4$ and $6$ satellites in collaboration is approximately $4.2$ dB and $4$ dB, respectively. However, when the quantization bit width is increased to $7$, the combined SNR loss for these two cases is reduced to $0.5$ dB and $0.4$ dB, respectively. It can be observed that increasing the quantization bit width improves the combined SNR, bringing it closer to its theoretical value. However, a higher quantization bit width also results in increased computational complexity. To strike a balance between combining performance and computational complexity, we choose a frequency offset quantization bit width of $6$ bits for the subsequent simulations in this paper. This choice provides a reasonable compromise between achieving satisfactory combining performance and managing computational complexity effectively.
	%补充D=6的SNRLoss，20230920完成以上部分修改
	Then, the relationship between the combined SNR loss and the number of iterations is simulated. Simulation parameters are set as $D = 6$, $E_ {\mathrm {s}} / N_ {\mathrm {0}} = -3$ dB, the number of satellites $M = 4$, the range of CFO of received signal at each satellite ($-7.8125\times10^{-3}$, $+7.8125\times10^{-3}$]. As can be seen from Fig. \ref{Ne}, when $N_{\mathrm{e}}=4$, it takes $4$ iterations to converge, and the combined SNR loss is about $1.3$ dB. when $N_{\mathrm{e}}=32$, it takes $8$ iterations to converge, and the combined SNR loss is only $0.4$ dB. This is because when the number of optimal combinations is small, although the convergence rate is faster, it is possible to obtain the local optimal solution. With the increase of $N_{\mathrm{e}}$, the global optimal solution can be obtained and the small Combined SNR loss can be obtained. However, the number of iterations also increases, resulting in an increase in computation. To select the appropriate number of optimal combinations, it is necessary to ensure that the combined SNR loss converges to the global optimal solution under the condition of not increasing too much computation.
	
	\begin{figure}
		\centering
		\includegraphics[width=0.48\textwidth]{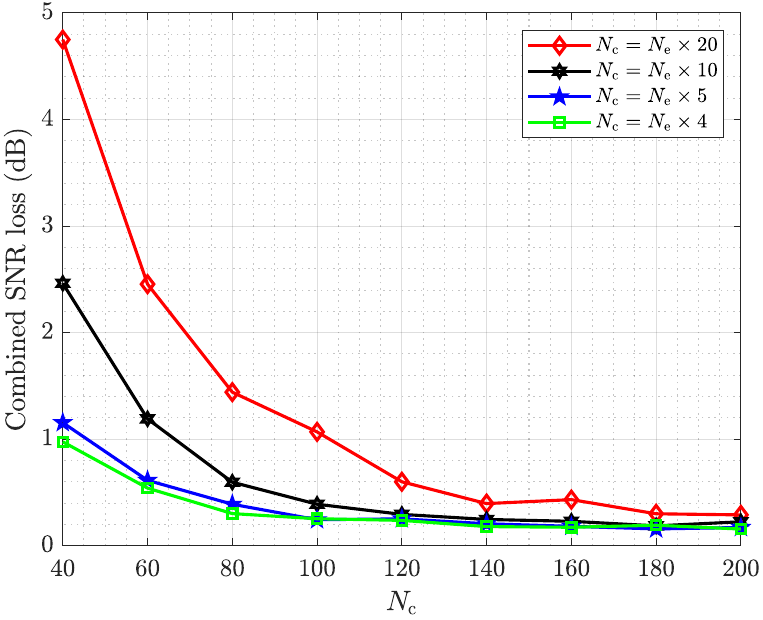}
		\caption{The relationship between $N_{\mathrm{c}}$ and $N_{\mathrm{e}}$ }\label{NcNe}
	\end{figure}
	
	Given $D=6$, $E_ {\mathrm {s}} /N_ {\mathrm {0}}=-3 $ dB, Fig.\ref {NcNe} shows the relationship between the number of random combinations and the combined SNR loss. It can be seen from Fig.\ref {NcNe} that the more combinations are combined in each iteration, the larger the range of both CFOs and CPOs at each iteration, and it is more possible to converge to the optimal value. However, when the number of random combinations reaches the $120$ group, the optimal value that converges tends to stabilize, and further increasing the number of random combinations will increase the complexity. Additionally, Fig.\ref {NcNe} also depicts the relationship between the number of random combinations and the number of optimal combinations. When the number of optimal combinations is small, it is easy to converge to the local optimal value. When the number of optimal signal combinations is $1 /5 $ of the number of random signal combinations, it is more possible to obtain the optimal solution.
	
	\subsection{Cooperative Expectation Maximization based Fine CFOs and CPOs Estimation}
	\subsubsection{Objective Function}

 In the initial coarse estimation stage, the uncertain range of CFO is reduced from $\left (-\frac{1}{2I},+\frac{1}{2I}\right ]$ to $\left (-\frac{1}{2^{D+1}I},+\frac{1}{2^{D+1}I}\right ]$. However, due to the quantization of frequency offsets, accurate estimation becomes impossible within the existing quantization interval. To address this limitation and improve estimation accuracy, we propose a Cooperative Expectation Maximization (CEM) method. It leverages the principles of the EM algorithm, which is widely used for estimating parameters in statistical models involving latent variables. The EM algorithm is particularly suitable for solving the MLE problem when direct maximization of the likelihood function is challenging \cite{EMCode}.

 In the fine estimation stage, the initial uncertain range of CFOs and CPOs is based on the residual set obtained from compensating the received signals using the estimation results from the ICE algorithm. This set is denoted as $\boldsymbol{\theta}_{\mathrm{res}}$. It represents the remaining CFOs and CPOs that need to be accurately estimated in the fine estimation stage, which can be written as:
	\begin{align}\label{ReceiverDitital77}
		\boldsymbol{\theta}_{\mathrm{res}} &= [f_{\mathrm{res},1},\phi_{\mathrm{res},1};\cdots;f_{\mathrm{res},m},\phi_{\mathrm{res},m};\cdots;f_{\mathrm{res},M},\phi_{\mathrm{res},M}] \notag\\
		&= [\boldsymbol{\Phi}_{1};\boldsymbol{\Phi}_{2};\boldsymbol{\Phi}_{m};\cdots;\boldsymbol{\Phi}_{M}].
	\end{align}
	where $\boldsymbol{\Phi}_{m} = [f_{\mathrm{res},m},\phi_{\mathrm{res},m}]$ represents the residual CFO and CPO to be estimated for the $m$-th satellite. Define the $k$-th compensated received symbol of the $m$-th satellite as $r_{m,k}$. Given the AWGN channel, the joint density function of $r_{m,k}$ is expressed as \eqref{pdfMLE}. 
	\begin{figure*}[ht]
		\centering
		\begin{align}\label{pdfMLE}
			p\left(r_{m,k} \mid \boldsymbol{\Phi}_{m}, s_{k}\right)&=\frac{1}{\pi N_{m}}\exp \left\{-\frac{|r_{m, k}- s^{*}_{k}\mathrm{e}^{j(2 \pi k f_{\mathrm{res},m} T_{\mathrm{s}}+\phi_{\mathrm{res},m})}|^2}{N_{m}}\right\} \notag\\
		&=\frac{1}{\pi N_{m}}\exp \left\{-\frac{|r_{m, k}|^2 - 2 \Re \left[s_{k}^{*}\mathrm{e}^{-j(2 \pi k f_{\mathrm{res},m} T_{\mathrm{s}}+\phi_{\mathrm{res},m})} r_{m, k}\right] + |s_{k}|^2}{N_{m}}\right\}
		\end{align}
		\rule{18cm}{0.01cm}		
	\end{figure*}
    Following the same approach as \eqref{pdf2}, the influence of the transmitted data $s_{k}$ can be removed from the joint density function in \eqref{pdfMLE}. The resulting conditional probability density function is solely dependent on the parameters $\boldsymbol{\Phi}_{m}$. 
    Using this simplified representation, \eqref{pdfMLE} can be updated as follows:
	\begin{align}\label{pdfMLE11}
		p\left(r_{m,k} \mid \boldsymbol{\Phi}_{m} \right)&=\mathbb{E}_{\boldsymbol{s}}\left[p_{k}(\mathfrak{m}) p\left(r_{m,k} \mid f_{\mathrm{res},m}, \phi_{\mathrm{res},m}, s_{k}\right)\right] \notag\\
		&=\sum_{s_{k} \in \mathfrak{M}} p_{k}(\mathfrak{m}) p\left(r_{m,k} \mid f_{\mathrm{res},m}, \phi_{\mathrm{res},m}\right).	
	\end{align}
     where, $\mathfrak{m}$ is the order of modulation. Let the $\boldsymbol{r}_{m}$ represent the all the received $K$ symbols of the $m$-th satellite, due to the independence of $r_{m,k}$, $\forall k = 0,1,\cdots,K-1$, the conditional probability density function of $\boldsymbol{r}_{m}$ is written as $p(\boldsymbol{r}_{m}\mid \boldsymbol{\Phi}_{m}) = \prod\limits_{k=0}^{K-1}p(r_{m,k} \mid \boldsymbol{\Phi}_{m})$.
	%提一下进制数
	By removing the constant terms that do not affect the parameter estimation, the logarithm of $p(\boldsymbol{r}_{m}\mid \boldsymbol{\Phi}_{m})$ can be given by,
	\begin{align}\label{pdfMLElog}
		\ln p\left(\boldsymbol{r}_{m} \mid \boldsymbol{\Phi}_{m} \right)=&\sum_{k=0}^{K-1}\ln \bigg\{\sum_{s_{k} \in \mathfrak{M}} p_{k}(\mathfrak{m})\notag\\
		&\times \mathrm{e}^{\Re \left[s_{k}^{*}\mathrm{e}^{-j(2 \pi k f_{\mathrm{res},m} T_{\mathrm{s}} + \phi_{\mathrm{res},m})} r_{m, k}\right]}\bigg\}.
	\end{align}
 
    It is a well-known fact that when $x$ is small, approaching zero, certain approximations can be made. Specifically, $\exp(x) \approx 1 + x$, $\ln(1 + x)\approx x$ \cite{Iterative2020}. In the communication scenarios being considered, the received signal $r_{m, k}$ exhibits low values. As a result, based on this information, \eqref{pdfMLElog} can be expressed in the following form:
	\begin{align}\label{pdfMLElog1}
		\ln p\left(\boldsymbol{r}_{m} \mid \boldsymbol{\Phi}_{m} \right)=&\sum_{k=0}^{K-1}\ln \bigg\{\sum_{s_{k} \in \mathfrak{M}} p_{k}(\mathfrak{m})\notag\\
		&\times\mathrm{e}^{\Re \left[s_{k}^{*}\mathrm{e}^{-j(2 \pi k f_{\mathrm{res},m} T_{\mathrm{s}}+ \phi_{\mathrm{res},m})} r_{m, k}\right]}\bigg\} \notag\\
%		\approx& \sum_{k=0}^{K-1}\ln \bigg\{\sum_{s_{k} \in \mathfrak{M}} p_{k}(\mathfrak{m}) \notag\\ &\times \bigg\{ 1+\Re \left[s_{k}^{*}r_{m,k}\mathrm{e}^{-j(2 \pi k f_{\mathrm{res},m} T_{\mathrm{s}} + \phi_{\mathrm{res},m})} \right]\bigg\} \bigg\} \notag\\
		\approx& \Re\bigg\{\sum_{k=0}^{K-1}\sum_{s_{k} \in \mathfrak{M}} p_{k}(\mathfrak{m})\notag\\
		&\times s_{k}^{*}r_{m,k}\mathrm{e}^{-j(2 \pi k f_{\mathrm{res},m} T_{\mathrm{s}} + \phi_{\mathrm{res},m})} \bigg\} \notag\\
		\approx& \Re\left\{\sum_{k=0}^{K-1}\eta_{k}^{*}r_{m,k}\mathrm{e}^{-j(2 \pi k f_{\mathrm{res},m} T_{\mathrm{s}} + \phi_{\mathrm{res},m})} \right\},
	\end{align}
	where $\eta_{k} = \sum\limits_{s_{k} \in \mathfrak{M}} p_{k}(\mathfrak{m})s_{k}$ represents the expectation of the transmitted data. However, determining the prior probability distribution of the transmitted data is generally not feasible or unknown, which renders the accurate calculation of $\eta_{k}$ difficult. Therefore, alternative approaches must be employed to estimate this parameter.
 
 \subsubsection{Algorithm Design}
    In this paper, we utilize the EM algorithm to address the aforementioned problem. The EM algorithm follows a specific framework that involves defining certain parameters of interest, observed signals, hidden signals. Specifically, for each satellite indexed by $m$, we define the parameter of interest as  $\boldsymbol{\Phi}_{m} = [f_{\mathrm{res},m},\phi_{\mathrm{res},m}]$. Additionally, we consider the observed signal $\boldsymbol{r}_{m}$ and the hidden signal, which corresponds to the transmitted data denoted by $\boldsymbol{s}$. To facilitate the EM algorithm, we define the complete observed signal $\mathbf{z}_{m} = [\boldsymbol{r}_{m}, \boldsymbol{s}]$ as the combination of the observed signal $\boldsymbol{r}_{m}$ and the transmitted data $\boldsymbol{s}$ for the $m$-th satellite. This allows us to incorporate both observed and hidden information in our analysis. The EM algorithm operates through two main steps that alternate iteratively: the E-step and the M-step. These steps enable the estimation of parameters by iteratively updating their values based on the available observed and hidden information. The algorithm proceeds by iteratively performing these steps until convergence is achieved, providing estimates for the desired parameters. The steps of the EM algorithm are as follows:
	
	$\mathrm{E}-\mathrm{step}$:
	\begin{align}\label{Estep}
		\mathcal{Q}(\boldsymbol{\Phi}_{m}; \boldsymbol{\Phi}_{m}^{(n-1)}) &= \mathbb{E}_{\mathbf{z}_{m} \mid \boldsymbol{r}_{m}, \boldsymbol{\Phi}_{m}^{(n-1)}}[\ln p(\mathbf{z}_{m} \mid \boldsymbol{\Phi}_{m})] \notag\\
		&=\int_{\boldsymbol{s}} p\left(\boldsymbol{s} \mid \boldsymbol{r}_{m}, \boldsymbol{\Phi}_{m}^{(n-1)}\right) \ln p(\mathbf{z}_{m} \mid \boldsymbol{\Phi}_{m}) d \boldsymbol{s} 
	\end{align}
	
	$\mathrm{M}-\mathrm{step}$:
	\begin{align}\label{Mstep}
		\boldsymbol{\Phi}_{m}^{(n)}= \arg \max_{\boldsymbol{\Phi}_{m}} \mathcal{Q}(\boldsymbol{\Phi}_{m}; \boldsymbol{\Phi}_{m}^{(n-1)}),
	\end{align}
In each iteration $n$ of the EM algorithm, we have an E-step and an M-step. The E-step involves calculating the expectation $\mathbb{E}_{\mathbf{z}_{m} \mid \boldsymbol{r}_{m}, \boldsymbol{\Phi}_{m}^{(n-1)}}[\ln p(\mathbf{z}_{m} \mid \boldsymbol{\Phi}_{m})]$, which represents the conditional probability distribution $p\left(\boldsymbol{s} \mid \boldsymbol{r}_{m}, \boldsymbol{\Phi}_{m}^{(n-1)}\right)$ of the hidden data $\boldsymbol{s}$ given the observed signal and the parameter of interest $\boldsymbol{\Phi}_{m}$. This expectation is calculated by evaluating the logarithm of the likelihood function $\ln p(\mathbf{z}_{m} \mid \boldsymbol{\Phi}_{m})$ can be given. $\mathrm{M}-\mathrm{step}$. After the E-step, we move to the M-step, where we aim to maximize the quantity $\mathcal{Q}(\boldsymbol{\Phi}_{m}; \boldsymbol{\Phi}_{m}^{(n-1)})$ with respect to the parameter of interest $\boldsymbol{\Phi}_{m}$. This maximization step provides an estimate for the parameter $\boldsymbol{\Phi}_{m}$, which then serves as the updated parameter for estimating $\boldsymbol{\Phi}_{m}^{(n)}$ in the next iteration. The iterative process of performing the E-step and M-step continues until the algorithm converges, meaning that further iterations do not significantly improve the estimates. At convergence, we obtain the MLE results for the CFOs and CPOs of the system. Since the transmitted data $\boldsymbol{s}$ and the parameter of interest $\boldsymbol{\Phi}_{m}$ are mutually independent, the quantity $\mathcal{Q}(\boldsymbol{\Phi}_{m}; \boldsymbol{\Phi}_{m}^{(n-1)})$ can be expressed as:
	\begin{align}\label{Estep1}
		\mathcal{Q}(\boldsymbol{\Phi}_{m}; \boldsymbol{\Phi}_{m}^{(n-1)}) & = \int_{\boldsymbol{s}} p\left(\boldsymbol{s} \mid \boldsymbol{r}_{m}, \boldsymbol{\Phi}_{m}^{(n-1)}\right) \ln p(\mathbf{z}_{m} \mid \boldsymbol{\Phi}_{m}) d \boldsymbol{s} \notag \\
		& = \int_{\boldsymbol{s}} p\left(\boldsymbol{s} \mid \boldsymbol{r}_{m}, \boldsymbol{\Phi}_{m}^{(n-1)}\right) \ln p(\mathbf{r}_{m},\boldsymbol{s} \mid \boldsymbol{\Phi}_{m}) d \boldsymbol{s} \notag\\
		& = \int_{\boldsymbol{s}} p\left(\boldsymbol{s} \mid \boldsymbol{r}_{m}, \boldsymbol{\Phi}_{m}^{(n-1)}\right) \ln p(\mathbf{r}_{m} \mid \boldsymbol{\Phi}_{m},\boldsymbol{s}) d \boldsymbol{s} \notag\\ &+ \int_{\boldsymbol{s}} p\left(\boldsymbol{s} \mid \boldsymbol{r}_{m}, \boldsymbol{\Phi}_{m}^{(n-1)}\right) \ln p(\boldsymbol{s}) d \boldsymbol{s}.
	\end{align}
 
	The right-hand side of \eqref{Estep1} is independent of the parameter $\boldsymbol{\Phi}_{m}$, hence $\mathcal{Q}(\boldsymbol{\Phi}_{m}; \boldsymbol{\Phi}_{m}^{(n-1)})$ can be simplified as:
	\begin{equation}\label{Estep2}
		\begin{aligned}
			\mathcal{Q}(\boldsymbol{\Phi}_{m}; \boldsymbol{\Phi}_{m}^{(n-1)}) & = \int_{\boldsymbol{s}} p\left(\boldsymbol{s} \mid \boldsymbol{r}_{m}, \boldsymbol{\Phi}_{m}^{(n-1)}\right) \ln p(\mathbf{r}_{m} \mid \boldsymbol{\Phi}_{m},\boldsymbol{s}) d \boldsymbol{s}
		\end{aligned}
	\end{equation}
     By substituting \eqref{pdfMLE} into \eqref{Estep2}, $\mathcal{Q}(\boldsymbol{\Phi}_{m}; \boldsymbol{\Phi}_{m}^{(n-1)})$ can be updated as \eqref{Estep3}.
	\begin{figure*}[ht]
		\centering
		\begin{equation}\label{Estep3}
			\begin{aligned}
				\mathcal{Q}(\boldsymbol{\Phi}_{m}; \boldsymbol{\Phi}_{m}^{(n-1)}) & =\int_{\boldsymbol{s}} p\left(\boldsymbol{s} \mid \boldsymbol{r}_{m}, \boldsymbol{\Phi}_{m}^{(n-1)}\right) \sum_{k=0}^{K-1}\left(\Re \left\{r_{m,k} s_{k}^{*} \mathrm{e}^{-j(2 \pi k f_{\mathrm{res},m} T_{\mathrm{s}} + \phi_{\mathrm{res},m})}\right\}-\left|s_{k}\right|^{2}\right) d \boldsymbol{s} \\
				& =\sum_{s_{k} \in \mathfrak{M}} p\left(s_{k} \mid \boldsymbol{r}_{m}, \boldsymbol{\Phi}_{m}^{(n-1)}\right) \sum_{k=0}^{K-1} \Re \left\{r_{m,k} s_{k}^{*} \mathrm{e}^{-j(2 \pi k f_{\mathrm{res},m} T_{\mathrm{s}} + \phi_{\mathrm{res},m})}\right\} \\
				& =\Re \left\{\sum_{k=0}^{K-1} r_{m,k}\left(\sum_{s_{k} \in \mathfrak{M}} s_{k}^{*} p\left(s_{k} \mid \boldsymbol{r}_{m}, \boldsymbol{\Phi}_{m}^{(n-1)}\right)\right) \mathrm{e}^{-j(2 \pi k f_{\mathrm{res},m} T_{\mathrm{s}} + \phi_{\mathrm{res},m})}\right\}
			\end{aligned}
		\end{equation}
		\rule{18cm}{0.01cm}
	\end{figure*}	
 
 Let $\zeta_{k}(\boldsymbol{r}_{m}, \boldsymbol{\Phi}_{m}^{(n-1)}) = \sum\limits_{s_{k} \in \mathfrak{M}} s_{k} p\left(s_{k} \mid \boldsymbol{r}_{m}, \boldsymbol{\Phi}_{m}^{(n-1)}\right)$ denotes the posterior expectation of the transmitted data given the received signal $\boldsymbol{r}_{m}$ and the parameter estimation $\boldsymbol{\Phi}_{m}^{(n-1)}$. Since each satellite in the system receives the same transmitted data, we can update the posterior expectation of the combined received signals at all $M$ satellites as:
	\begin{align}\label{Estep4}
		\zeta_{k}(\boldsymbol{r}_{\mathrm{com}}, \boldsymbol{\Phi}^{(n-1)})  =& \sum\limits_{s_{k} \in \mathfrak{M}} s_{k} p\left(s_{k} \mid \boldsymbol{r}_{\mathrm{com}}, \boldsymbol{\Phi}^{(n-1)}\right) \notag\\
		=& \sum\limits_{s_{k} \in \mathfrak{M}} s_{k} p\bigg(s_{k} \mid \sum \limits_{m=1}^{M}\sum \limits_{k=0}^{K-1} r_{m, k} \notag\\
		&\times \mathrm{e}^{-j(2 \pi k f_{\mathrm{res},m}^{(n-1)} T_{\mathrm{s}} + \phi_{\mathrm{res},m}^{(n-1)})}, \boldsymbol{\Phi}^{(n-1)}\bigg)
	\end{align}
	where $f_{\mathrm{res},m}^{(n-1)}$ and $\phi_{\mathrm{res},m}^{(n-1)}$ represents the residual CFO and CPO of the received signal at the $m$-th satellite in the $(n-1)$-th iteration, respectively. Equation \eqref{Estep4} captures the combination gain achieved through multi-satellite cooperation in the parameter estimation process, which characterizes the algorithm as Cooperative EM. It reflects how the posterior expectation of the transmitted symbols is calculated by utilizing the combined results from all satellites. By leveraging information from multiple receivers, the resulting expectation derived from this collective fusion of results exhibits improved reliability compared to individual satellite-based calculations. Therefore,  $\mathcal{Q}(\boldsymbol{\Phi}_{m}; \boldsymbol{\Phi}_{m}^{(n-1)})$ can be updated as follows:
	\begin{align}\label{Estep5}
		\mathcal{Q}(\boldsymbol{\Phi}_{m}; \boldsymbol{\Phi}_{m}^{(n-1)})  =&\Re \bigg\{\sum_{k=0}^{K-1} r_{m,k}\zeta_{k}^{*}(\boldsymbol{r}_{\mathrm{com}}, \boldsymbol{\Phi}^{(n-1)}) \notag\\
		&\times \mathrm{e}^{-j(2 \pi k f_{\mathrm{res},m} T_{\mathrm{s}} + \phi_{\mathrm{res},m})}\bigg\}.
	\end{align}
 
	The expression $\mathcal{Q}(\boldsymbol{\Phi}_{m}; \boldsymbol{\Phi}_{m}^{(n-1)})$ shares the same mathematical form as \eqref{pdfMLElog1}. However, there exists a key distinction is that $\eta_{k}$ is the prior expectation of the transmitted symbols, while $\zeta_{k}(\boldsymbol{r}_{\mathrm{com}}, \boldsymbol{\Phi}^{(n-1)})$ is the posterior expectation of the transmitted symbols obtained from the combined results of $M$ satellites. The determination of this posterior expectation involves computing the posterior LLR information based on the combined decoding outcomes. The $\mathrm{M-step}$ for finding the maximum value in an array becomes:
	\begin{align}
		\boldsymbol{\Phi}_{m}^{(n)} =& \arg \max_{\boldsymbol{\Phi}_{m}} \Re \bigg\{\sum_{k=0}^{K-1} r_{m,k}\zeta_{k}^{*}(\boldsymbol{r}_{\mathrm{com}}, \boldsymbol{\Phi}^{(n-1)})\notag\\ 
		&\times\mathrm{e}^{-j(2 \pi k f_{\mathrm{res},m}^{(n-1)} T_{\mathrm{s}} + \phi_{\mathrm{res},m}^{(n-1)})}\bigg\}.
	\end{align}

 Accordingly, we can express the iterative process of the residual CFO and CPO of the received signal at the $m$-the satellite as:
	\begin{equation}\label{Mstep2}
		\begin{aligned}
			f_{\mathrm{res},m}^{(n)}&=\arg \max\limits_{f_{\mathrm{res},m}}\left|\sum\limits_{k=0}^{K-1} r_{m,k} \zeta_{k}^{*}(\boldsymbol{r}_{\mathrm{com}}, \boldsymbol{\Phi}^{(n-1)}) \mathrm{e}^{-j 2 \pi k f_{\mathrm{res},m} T_{\mathrm{s}}}\right| \notag\\
		\end{aligned}
	\end{equation}
	\begin{equation}\label{Mstep3}
		\begin{aligned}
			\phi_{\mathrm{res},m}^{(n)}&=\arg \left\{\sum\limits_{k=0}^{K-1} r_{m,k} \zeta_{k}^{*}(\boldsymbol{r}_{\mathrm{com}}, \boldsymbol{\Phi}^{(n-1)}) \mathrm{e}^{-j 2 \pi k f_{\mathrm{res},m}^{(n)} T_{\mathrm{s}}}\right\}
		\end{aligned},
	\end{equation}
	\eqref{Mstep2} can be solved by quantizing the frequency within a specific range and searching for the maximum value to obtain $f_{\mathrm{res},m}^{(n)}$. As aforementioned, in the ICE algorithm output, the range of CFO has been narrowed down from $\left (-\frac{1}{2I},+\frac{1}{2I}\right ]$ to $\left (-\frac{1}{2^{D+1}I},+\frac{1}{2^{D+1}I}\right ]$, where $D$ represents the number of bits used for quantization. In this fine estimation process, each satellite utilizes the EM algorithm to perform a search for the residual frequency offset within the range $\left (-\frac{1}{2^{D+1}I},+\frac{1}{2^{D+1}I}\right ]$. In this scenario, a search step of $f_{\mathrm{step}}$ is employed to calculate the combined results. Consequently, the residual CFO and CPO can be updated through an iterative as:
	\begin{equation}\label{Mstep5}
		\begin{aligned}
			\boldsymbol{r}_{\mathrm{com}}^{(n-1)} &= \sum \limits_{m=1}^{M}\sum \limits_{k=0}^{K-1} r_{m, k}^{(n-1)} \mathrm{e}^{-j(2 \pi k f_{\mathrm{res},m}^{(n-1)} T_{\mathrm{s}} + \phi_{\mathrm{res},m}^{(n-1)})} \\
		\end{aligned}
	\end{equation}
	\begin{equation}\label{Mstep6}
		\begin{aligned}
			f_{\mathrm{res},m}^{(n)}&=\arg \max\limits_{f_{\mathrm{res},m}}\left|\sum\limits_{k=0}^{K-1} r_{m,k}^{(n-1)} \zeta_{k}^{*}(\boldsymbol{r}^{(n-1)}_{\mathrm{com}}, \boldsymbol{\Phi}^{(n)}) \mathrm{e}^{-j 2 \pi k f_{\mathrm{step}} T_{\mathrm{s}}}\right| \\
		\end{aligned}
	\end{equation}
	\begin{equation}\label{Mstep7}
		\begin{aligned}
			\phi_{\mathrm{res},m}^{(n)}&=\arg \left\{\sum\limits_{k=0}^{K-1} r_{m,k} \zeta_{k}^{*}(\boldsymbol{r}_{\mathrm{com}}^{(n-1)}, \boldsymbol{\Phi}^{(n-1)}) \mathrm{e}^{-j 2 \pi k f_{\mathrm{res},m}^{(n)} T_{\mathrm{s}}}\right\}.
		\end{aligned}
	\end{equation}

    During each iteration of the CEM algorithm, the estimated results are utilized to compensate for the received signal of each satellite. This compensation enables coherent combination and decoding of the signals. Additionally, the posterior LLR information obtained from the decoder is employed to calculate the expected value of the transmitted data. As the number of iterations increases, the estimated values of the CFOs and CPOs gradually approach the actual values. Simultaneously, the reliability of the posterior LLR information from the decoded output after coherent combination improves.

    In summary, the joint utilization of the Iterative Cross Entropy (ICE) algorithm and the Cooperative Expectation Maximization (CEM) algorithm, referred to as ICE-CEM algorithm, enables accurate estimation of CFOs and CPOs within the range of $\left(-\frac{1}{2I},+\frac{1}{2I}\right]$ and $(-\pi,+\pi]$. This accurate estimation facilitates the completion of coherent combination among cooperative satellites. The flow of the ICE-CEM algorithm is presented in Algorithm 2.
	
	\begin{algorithm}
		\SetAlgoLined
		\KwIn{Received signal at all $M$ satellites $\mathbf{r} = [\boldsymbol{r}_{1},\boldsymbol{r}_{2},\cdots,\boldsymbol{r}_{M}]^{\mathrm{T}}$}
		\KwOut{Estimated frequency and phase offsets for the despreaded data of the $m$-th satellite $\boldsymbol{\hat{\Phi}} = [\Delta \hat{f}_{1},\cdots, \Delta \hat{f}_{m},\Delta \hat{f}_{M}, \hat{\phi}_{1},\cdots,\hat{\phi}_{m},\hat{\phi}_{M}]$ and the decoded result after coherent combination, $\Delta \hat{f}_{m} = \Delta \hat{f}_{m,D} + \hat{f}_{\mathrm{res},m}$; $\hat{\phi}_{m} = \hat{\phi}_{m,D} +\hat{\phi}_{\mathrm{res},m}$, with the decoded symbols denoted as $\hat{\boldsymbol{u}}.$}	
		Initialization: ICE algorithm parameters $N_{\mathrm{c}}$ (initial candidate group number), $N_{\mathrm{e}}$ (optimal group number), $N_{\mathrm{iter}}$ (maximum number of iterations); $N_{\mathrm{iter,EM}}$ (maximum number of iterations);\\
		(1) Apply the ICE algorithm to obtain the estimated results of frequency offset and phase offset for the decoded symbols $\Delta \hat{f}_{m,D}$ and $\hat{\phi}_{m,D}$ of each satellite; Compensate the despreaded signal of each satellite using $\Delta \hat{f}_{m,D}$ and $\hat{\phi}_{m,D}$;\\
		\For{$1 \leq n \leq N_{\mathrm{iter,EM}}$}{
			(2)Merge the compensated results for each satellite using \eqref{Mstep5}. Perform Polar code decoding and obtain the posterior LLR information. Calculate the mathematical expectation $\zeta_{k}(\boldsymbol{r}_{\mathrm{com}}, \boldsymbol{\Phi}^{(n-1)})$  of the transmitted symbols using equation\eqref{Lpos};\\
			(3) Execute \eqref{Mstep6} and \eqref{Mstep7} sequentially to obtain accurate estimations of residual frequency offset and phase offset for the decoded symbols of each satellite;\\
			(4) $n = n + 1$;\\
		}
		return:$\boldsymbol{\hat{\Phi}}$, the decoded data $\hat{\boldsymbol{u}}$
		\caption{ICE-CEM algorithm}
		\label{euclid2}
	\end{algorithm}

	\section{Simulation and Performance Analysis}
\begin{figure*}[ht]
		\centering
		\subfigure [RMSE performance of CFO estimation versus $E_{\mathrm{s}}/N_{\mathrm{0}}$]{
			\includegraphics [width=0.48\textwidth]{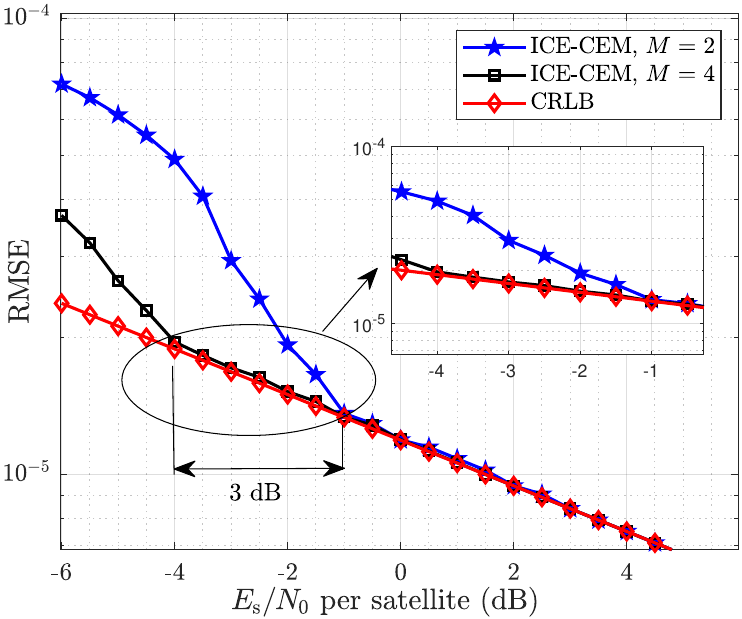}}
		\subfigure [RMSE performance of CPO estimation versus $E_{\mathrm{s}}/N_{\mathrm{0}}$]{
			\includegraphics [width=0.48\textwidth]{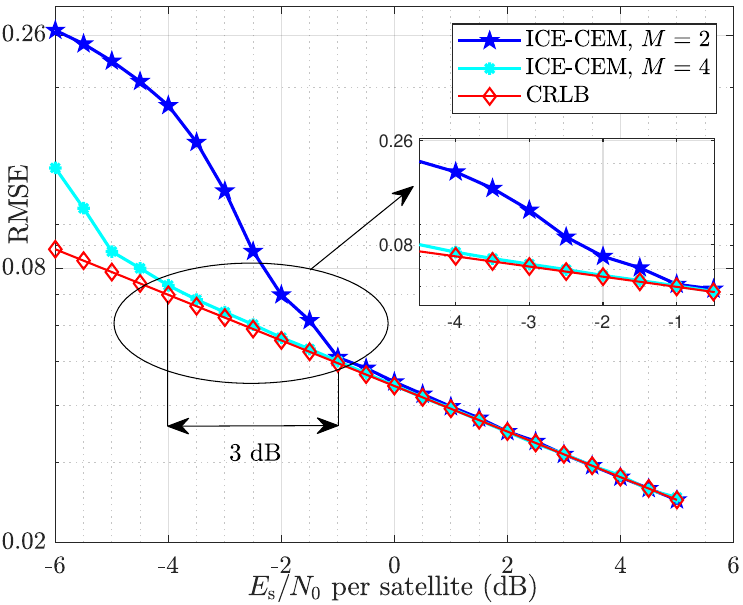}}
		\caption{RMSE performance in CFO and CPO estimation via the ICE-CEM algorithm when $M=2$ and $M=4$.}
		\label{RMSEPic}
	\end{figure*}

    In order to assess the performance of the proposed ICE-CEM algorithm in estimating CFO and CPO, we perform  Root Mean Square Error (RMSE) simulations and compare the results with Cramér-Rao Lower Bound (CRLB),  which represents the theoretical limits. Fig. \ref{RMSEPic} presents the performance of the ICE-CEM algorithm for cooperative scenarios involving 2 satellites and 4 satellites. Fig. \ref{RMSEPic}(a) and Fig. \ref{RMSEPic}(b) demonstrate that, for a fixed number of cooperative satellites, the estimated RMSE of CFO and CPO using the ICE-CEM algorithm closely approximate the CRLB as the SNR increases. This can be attributed to the fact that higher SNR leads to greater reliability of the posterior LLR information and more accurate calculation of $\zeta_{k}$ using \eqref{Lpos}. Furthermore, as the number of cooperative satellites increases from 2 to 4, the required SNR for the ICE-CEM algorithm to approximate the CRLB decreases approximately 3 dB. This indicates the cooperative gain achieved by the ICE-CEM algorithm. With more cooperative satellites, the coherent combined SNR increases, resulting in improved reliability of $\zeta_{k}$ after coherent combination and superior RMSE performance.

	\begin{figure*}[ht]
		\centering
		\subfigure [RMSE performance of CFO estimation in the range of $(-7.8125\times10^{-3}$, + $7.8125\times10^{-3})$ ]{
			\includegraphics [width=0.48\textwidth]{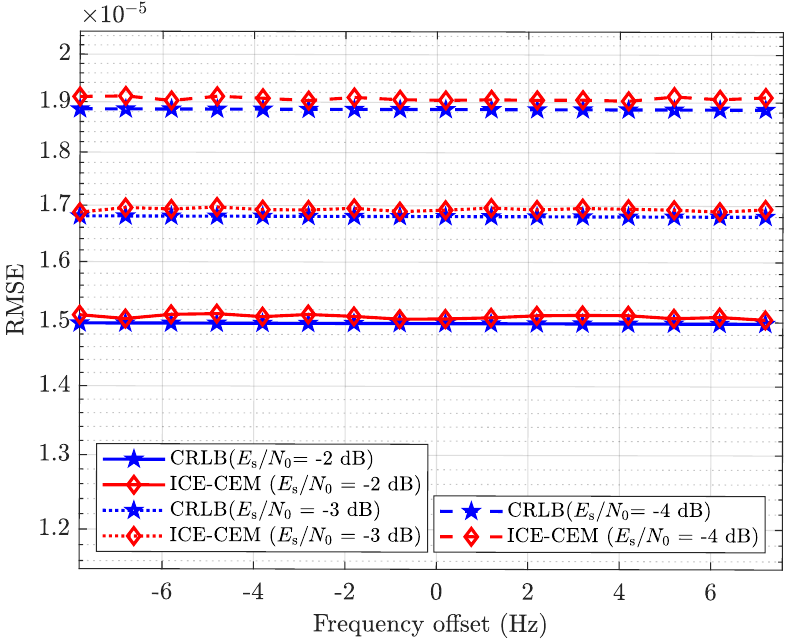}}
		\subfigure [RMSE performance of CPO estimation in the range of $(-\pi,+\pi)$]{
			\includegraphics [width=0.48\textwidth]{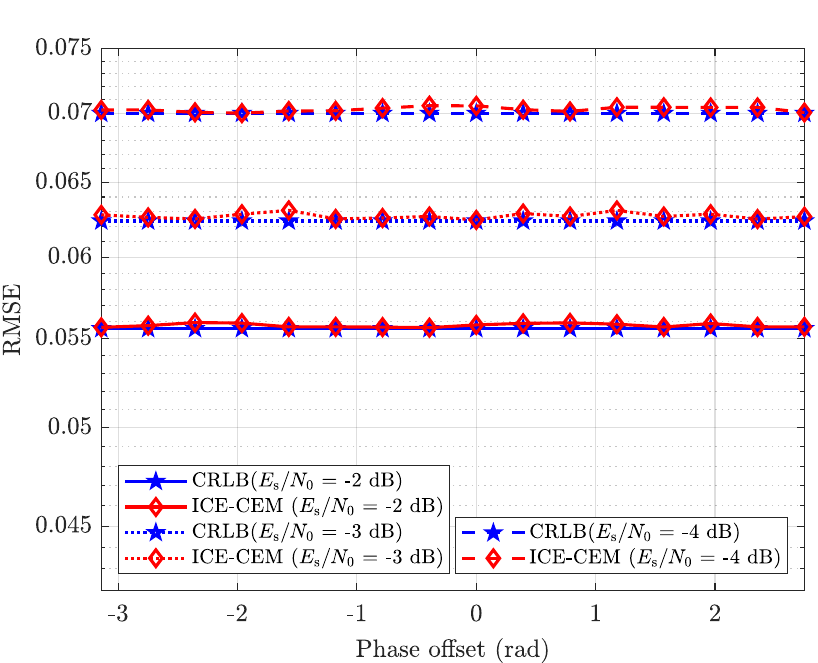}}
		\caption{RMSE performance in CFO and CPO estimation via ICE-CEM algorithm at varying $E_{\mathrm{s}}/N_{\mathrm{0}}$ ($M=4$).}
		\label{FrePhiSearch}
	\end{figure*}

Fig.~\ref{FrePhiSearch} demonstrates the estimation performance of the ICE-CEM algorithm for different ranges of CFO and CPO, considering various values of $E_{\mathrm{s}}/N_{\mathrm{0}}$. Subfigure \ref{FrePhiSearch}(a) illustrates the RMSE performance of NFO estimation within the range of $-7.8125\times10^{-3}$, + $7.8125\times10^{-3}$]. As $E_{\mathrm{s}}/N_{\mathrm{0}}$ increases, the RMSE gradually approaches the CRLB across the entire NFO range. Similarly, subfigure \ref{FrePhiSearch}(b) presents the RMSE performance of CPO estimation within the range of $(-\pi,+\pi]$. Again, as $E_{\mathrm{s}}/N_{\mathrm{0}}$ increases, the RMSE of the estimation results approaches the CRLB within the entire range.

\begin{figure}[ht]
		\centering
		\includegraphics[width=0.48\textwidth]{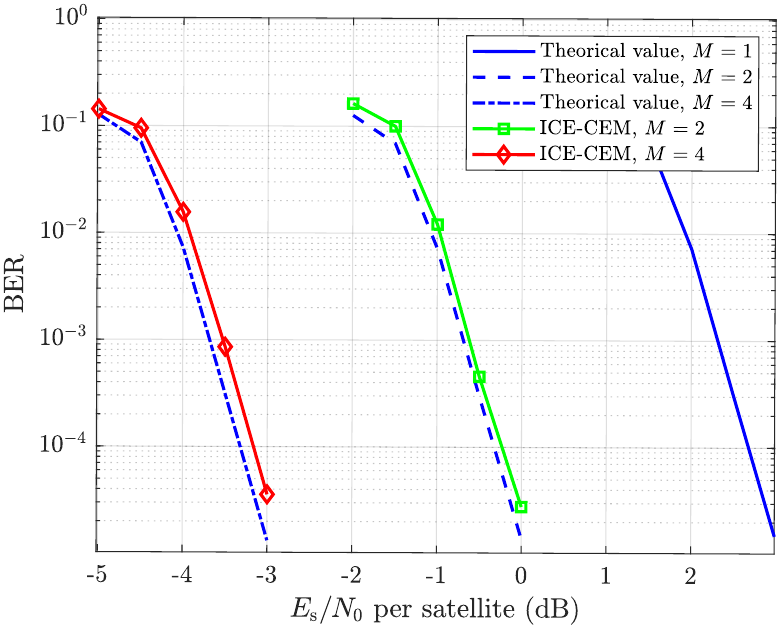}
		\caption{Simulation results of BER for two-satellite and four-satellite cooperation}\label{MulSatBERSim}
	\end{figure}

Additionally, Fig.~\ref{MulSatBERSim} displays the BER simulation results for scenarios involving 2 and 4 satellites. In these simulations, the estimation of CFOs and CPOs for each satellite is completed and the results are combined to compensate and decode the received signals using the BP decoder \cite{Chaofan1}. The NFO of each satellite is randomly selected within the range of ($-7.8125\times10^{-3}$, $+7.8125\times10^{-3}$], while the CPO is randomly chosen from the range of ($-\pi $, $+\pi$]. It is assumed that the received SNR is identical for each satellite.
It can be seen from Fig. ~\ref{MulSatBERSim} that the BER performance of the ICE-CEM algorithm incurs a loss of about $0.4$ dB and $0.3$ dB  (@BER=$1\times10^{-4}$) compared to the ideal scenario when the number of cooperative satellites is 2 and 4, respectively.  In the 4-satellites scenario, where each satellite operates at a lower SNR, the increased RMSE in CFO and CPO estimation for each satellite contributes to a degraded coherent combination among the satellites. Consequently, this leads to a larger loss in the combining and decoding performance.
	
	\section{Conclusions}
    This paper proposes an iterative code-aided estimation algorithm for CFOs and CPOs in the application of CSC, called ICE-CEM. The algorithm specifically aims to overcome the challenges associated with a wide range of CFOs and CPOs, poor synchronization accuracy under low SNR, and the absence of training sequences. The proposed algorithm begins with an ICE process, which quantifies the frequency offset and random phase offset for the received signal at each satellite. By employing cross-entropy iteration, it enables parallel search of CFOs and CPOs. Subsequently, the algorithm incorporates the CEM iteration algorithm to achieve a accurate estimation of CFO and CPO at each satellite.

 Simulation results utilizing the RMSE metric demonstrate the exceptional performance of the ICE-CEM algorithm in terms of both estimation range and accuracy. Specifically, the proposed algorithm achieves estimation accuracy close to the CRLB within the frequency range of ($-7.8125\times10^{-3}$, $+7.8125\times10^{-3}$] and phase range of ($-\pi $, $+\pi$]. Moreover, BER simulations reveal that the ICE-CEM algorithm incurs a mere loss of 0.3 dB and 0.4 dB (@BER=$1\times10^{-4}$) in the 2-satellites and 4-satellites scenarios, respectively.
	
	\bibliographystyle{IEEEtran}
	\normalem
	\bibliography{ref0909}
\end{document}